\documentclass[11pt, a4paper]{article}

  \usepackage[french]{babel}
  \usepackage[latin1]{inputenc}
\usepackage{microtype}
\usepackage{url}
\usepackage[plainpages=false]{hyperref}

   \usepackage{setspace}

\usepackage{pifont}

  \usepackage{multicol} 
  \usepackage{amsmath}
  \usepackage{amssymb}
  \usepackage{amscd}
 \usepackage{bbold} 
  \usepackage{stmaryrd}
  \usepackage{fancyhdr}

  \usepackage{shadow}
  \usepackage[usenames,dvipsnames]{xcolor}  

  \usepackage{tikz} 
\usetikzlibrary{matrix}
  \usepackage{pgflibraryarrows}
  \usepackage{extarrows}

  \usepackage{fourier-orns} 

\usepackage[shortlabels]{enumitem}
\definecolor{gris}{rgb}{0.9,0.9,0.9}
\definecolor{tomate}{rgb}{0.8,0,0}
\definecolor{apple}{rgb}{0.402,0.5,0}
\definecolor{myred}{rgb}{0.7,0,0}
\definecolor{myblue}{rgb}{0,0,0.6}
\definecolor{mygreen}{rgb}{0,0.4,0.3}
\definecolor{bluemarine}{rgb}{0,0,0.3}
\definecolor{quasiblack}{rgb}{0.1,0.1,0.2}
\definecolor{quasiblack}{rgb}{0.1,0.1,0.2}
\definecolor{gray}{rgb}{0.18,0.18,0.18}
\definecolor{quasiinvisible}{rgb}{0.95,0.95,0.95}
\definecolor{bluegray}{rgb}{0.1,0.2,0.3}
\definecolor{verylightgray}{rgb}{0.7,0.7,0.7}
\definecolor{verylightgray2}{rgb}{0.92,0.92,0.92}
\definecolor{lightgray}{rgb}{0.6,0.6,0.6}

\makeatletter

\renewcommand\paragraph{\@startsection{paragraph}{4}{-1.08cm}
  {0.3cm \@plus -1ex \@minus -.2ex}
  {-0.5em}
  {\bfseries}*
}

\makeatother
\newcounter{pgh}
\renewcommand{\thepgh}{{\S \arabic{pgh}~}}
\newcommand{\pgh}[1]{\paragraph{\makebox[8.5mm]{\footnotesize\color{black!50}\thepgh}}
    {\normalsize\bf #1}\refstepcounter{pgh}}

\newcounter{listitem}
\renewcommand{\thelistitem}{(\alph{listitem})~}
\newcommand{\litem}{{\color{black!60}\thelistitem}\refstepcounter{listitem}}
\newcounter{listitemr}
\renewcommand{\thelistitemr}{(\roman{listitemr})~}
\newcommand{\litemr}{{\color{black!60}\thelistitemr}\refstepcounter{listitemr}}
\newcounter{listitema}
\def\endlitem{\setcounter{listitem}{1}}
\def\endlitemr{\setcounter{listitemr}{1}}

\newcounter{hypcounter}[section]
\AtBeginDocument{\setcounter{hypcounter}{1}}

\def\thehypcounter{\itshape{(\roman{hypcounter})}}

  \newcommand{\commentbox}[7]{\bigskipp\hspace{-0.1\textwidth}\fcolorbox{black}{#1}{
      \begin{minipage}{1.2\textwidth}
        \textcolor{#2}{\titrebox{#3}{#4}{#6}}\par\par\smallskipp
        \textcolor{#5}{#7}
      \end{minipage}}\bigskipp}
  
\newcounter{remarkcounter} 
  \newenvironment{pr}[1]{\begin{trivlist}\item[\hskip\labelsep{Proof:}]~#1}{\end {trivlist}}
\newenvironment{myproof}{\begin{pr}}{\hfill\small{$\Box$} \end{pr}}

  \newenvironment{begpr}[1]{\begin{trivlist}\item[\hskip\labelsep{\bfseries
           Proof\ldots?}]~#1}{\end {trivlist}}

\usepackage{amsthm}

\newtheorem{lemma}{Lemma}
\newtheorem{conjecture}{Conjecture}
\newtheorem{proposition}{Proposition}
\newtheorem{example}{Example}

\newtheoremstyle{example}
{3pt}
{3pt}
{}
{}
{\upshape}
{:}
{.5em}
{}

\DeclareMathOperator{\sign}{sign}

\DeclareMathOperator{\traj}{{\ensuremath {\cal T}}}
\DeclareMathOperator{\orpot}{{\ensuremath {\cal P}^{\ast}}}
\DeclareMathOperator{\orpott}{{\ensuremath {\cal P}}}

\DeclareMathOperator{\reps}{{\ensuremath {\cal R}^\ast}}

\DeclareMathOperator{\cat}{{\ensuremath \inherits}}
\DeclareMathOperator{\inherits}{{\ensuremath \ltimes}}
\DeclareMathOperator{\inheritss}{{{\ensuremath \inherits^{\hspace{-0.5ex}\ast}}}}

\newcommand{\Htraj}{{\ensuremath  H^{\traj}}}
\newcommand{\AH}{{\ensuremath  A^H}}

\newcommand{\Aplus}{{\ensuremath  A_{+}^{\traj}}}
\newcommand{\Amoins}{{\ensuremath  A_{-}^{\traj}}}
\newcommand{\pot}[1]{{\ensuremath \langle #1 \rangle}}
\newcommand{\set}[1]{{\ensuremath \lbrace #1 \rbrace}}
\newcommand{\iset}[1]{{\ensuremath \llbracket #1 \rrbracket}}
\newcommand{\isetr}[1]{{\ensuremath \rrbracket #1 \rrbracket}}
\newcommand{\isetl}[1]{{\ensuremath \llbracket #1 \llbracket}}
\newcommand{\iseto}[1]{{\ensuremath \rrbracket #1 \llbracket}}

\makeatletter
\def\@maketitle{%
  \begin{center}%
    {\LARGE \@title \par}%
    \vskip 0.5em%
    {\large
      \lineskip .5em%
      \begin{tabular}[t]{c}%
        \@date
      \end{tabular}\par}%
    \vskip 1em%
  \end{center}%
  \par
  \vskip 1.5em}
\makeatother

\author{Mathilde Noual$^{1,2}$}
\begin{document}

\bibliographystyle{plain}


  \def\AGIM{Universit\'e Joseph Fourier -- Grenoble 1, AGIM, CNRS FRE 3405,\\ 38706 La Tronche, France}
  \def\LIP{Universit\'e de Lyon, \'ENS-Lyon, LIP, CNRS UMR 5668, 69007 Lyon, France}
  \def\IBISC{Universit\'e d'\'Evry -- Val d'Essonne, IBISC, \'EA 4526, 91000 \'Evry, France}
  \def\IXXI{IXXI, Institut rh\^one-alpin des syst\`emes complexes, 69007 Lyon, France}
  \def\detailsMN{\texttt{mathilde.noual@ens-lyon.fr}, Laboratoire d'informatique
    du parall{\'e}lisme de l'{\'E}NS Lyon, 46 all{\'e}e d'Italie, 69364 Lyon
    Cedex 07, France, tel : +33-(0)4-26-23-38-04, fax: +33-(0)4-26-23-38-20.}
  \def\SS{Sylvain Sen\'e}
  \def\MN{Mathilde Noual}
  \def\JD{Jacques Demongeot}

  \newcommand{\af}[1]{$^{#1}$}
  \newcommand{\aff}[2]{$^{#1,#2}$}
  \newcommand{\afff}[3]{$^{#1,#2,#3}$}

  \newenvironment{affiliations}
                 {\noindent \rule{14cm}{1pt} \begin{center} \small \begin{itemize}}
                 {\end{itemize} \end{center} \vspace{-5pt} \rule{14cm}{1pt}}


\newcommand{\figref}[1]{Figure~\ref{#1}}
\renewcommand{\eqref}[1]{(\ref{#1})}

\renewcommand{\contentsname}{Table of contents}
\renewcommand{\refname}{Another name for my references} 

\def\Hligne{\noindent\rule{14cm}{1pt}}
\def\sommaire{\Hligne\tableofcontents\Hligne}

\renewcommand{\bigskip}{\vspace{0.8cm}} 
\renewcommand{\medskip}{\vspace{0.4cm}} 
\renewcommand{\smallskip}{\vspace{0.2cm}} 
\def\medskipp{\vspace{0.4cm} \noindent}
\def\bigskipp{\vspace{0.8cm} \noindent}
\def\smallskipp{\vspace{0.2cm} \noindent}

\newcommand{\titrebox}[3]{\centerline{{\large #1~{\sc #3}~#2}}}
\newcommand{\titreboxDEF}[1]{\titrebox{$\lhd$}{$\rhd$}{#1}}
\newcommand{\quest}[1]{\commentbox{Goldenrod!70}{CadetBlue}{$\lhd$}{$\rhd$}{CadetBlue}{question de
      recherche}{{\bf #1}}}
\newcommand{\ajout}[1]{\commentbox{LimeGreen}{RedViolet}{$\oplus$}{$\oplus$}{RedViolet}{a
    ajo{\^u}ter}{{\bf #1}}}
\newcommand{\ajoutfig}[1]{\commentbox{RedViolet!80}{Aquamarine}{$\oplus$}{$\oplus$}{gray}{a
    ajo{\^u}ter}{{\bf #1}}}
\newcommand{\remq}[1]{\commentbox{CadetBlue!60}{Magenta}{$\lhd$}{$\rhd$}{gray}{remarque}{{\bf
      #1}}}
\newcommand{\citebox}[1]{\commentbox{Aquamarine}{gray}{$\star$}{$\star$}{gray}{citation}{{\bf #1}}}
\def\dangers{\danger\danger\danger\danger\danger\danger}
\newcommand{\problem}[1]{\commentbox{Red!80}{gray}{\dangers}{\dangers}{gray}{probl{\`e}me}{\centerline{{\sc #1}}}}
\def\threestar{$\star\star\star$}
\newcommand{\sun}[1]{\commentbox{gray!80}{White}{\threestar}{\threestar}{White}{``sun says''}{{\bf
      #1}}}
\newcommand{\sunco}[1]{\colorbox{gray!80}{\textcolor{White}{{\bf \threestar~#1~\threestar}}}}
\newcommand{\co}[1]{\colorbox{LimeGreen}{\textcolor{RedViolet}{{\bf $\lhd$~#1~$\rhd$}}}}
\newcommand{\addex}[1][]{\textcolor{BrickRed}{{\large $\mathbb{EX}$}[~#1~]} }
\newcommand{\cit}{\textcolor{Aquamarine}{{\large {\sc cite}}} }
\newcommand{\cpx}[1][]{\textcolor{ForestGreen}{{\huge $\copyright$}[~#1]}}

    
\def\check{ \textcolor{red}{{\large \danger} }}
\def\bigcheck{ \textcolor{red}{{\Huge \danger} }}




\newcommand{\forlater}[1]{#1}

\def\V{{\ensuremath V^{\ast}}}
\def\ie{{\it i.e.}}
\def\cf{{\it cf}}
\def\us{update schedule}
\def\uss{update schedules}
\def\bs{block-sequential}
\def\bsus{block-sequential update schedule}
\def\bsuss{block-sequential update schedules}
\def\pus{periodic update schedule}
\def\puss{periodic update schedules}
\def\GT{transition graph}
\def\GTs{transition graphs}
\def\STS{state transition system}
\def\STSs{state transition systems}
\def\DS{dynamical system}
\def\DSs{dynamical systems}
\def\BAN{Boolean automata network}
\def\BANs{Boolean automata networks}
\def\ltf{local transition function}
\def\ltfs{local transition functions}

\def\GIG{{\sc gtg}}
\def\GGIG{{\sc gtg}${}^{\star}$}
\def\SIG{{\sc stg}}
\def\SSIG{{\sc stg}${}^{\star}$}

\def\nieme[#1]{ {\ensuremath {#1}^{\mbox{\footnotesize th}}}}



\def\QED{\hfill{\ensuremath\Box}}
\def\U{\ensuremath{\mathcal{U}}}
\def\nU{\ensuremath{\overline{\U}}}
\def\01{\ensuremath{\{0,1\}}}
\def\B{\ensuremath{\mathbb{B}}}
\def\potentials{\ensuremath{\mathbb{P}}}
\def\alltimes{\ensuremath{\mathbb{T}}}
\def\SSS{\ensuremath{\mathbb{S}}}
\def\Bn{\ensuremath{\B^n}}
\def\N{\ensuremath{\mathbb{N}}}
\def\ntk{\ensuremath{\mathcal{N}}}
\def\Z{\ensuremath{\mathbb{Z}}}
\newcommand{\NN}[1][n]{\ensuremath{\N/#1\N}}
\newcommand{\ZZ}[1][n]{\ensuremath{\Z/#1\Z}}
\def\iff{{\ensuremath \ \Leftrightarrow\ }}
\def\Lar{{\ensuremath \ \Leftarrow\ }}
\def\Rar{{\ensuremath \ \Rightarrow\ }}

\newcommand{\SB}[0]{\scalebox{0.8}[1]{\ensuremath{\mathbf{S\hspace{-0.3ex}B}}}}
\newcommand{\BS}[0]{\scalebox{0.8}[1]{\ensuremath{\mathbf{B\hspace{-0.3ex}S}}}}
\def\mod{\mbox{ mod }}
\renewcommand{\bar}[2]{\ensuremath{\overline{#1}^{\scriptstyle #2}}}

\newcommand{\fcn}[4]{\ensuremath{\left\{\begin{array}{rcl} #1 & \to & #2\\#3 & \mapsto & #4\end{array}\right.}}
\newcommand{\fcnn}[2]{\ensuremath{\begin{array}{rcl} #1 & \mapsto & #2\end{array}}}

\def\BSn{\ensuremath{\mbox{{\sc bs}}_n} }
\def\BSnH{\ensuremath{\widetilde{\mbox{{\sc bs}}}_n} }
\def\BSk[#1]{\ensuremath{\mbox{{\sc bs}}_{#1}} }
\def\BSnH{\ensuremath{\widetilde{\mbox{{\sc bs}}}_n} }
\def\BSkH[#1]{\ensuremath{\widetilde{\mbox{{\sc bs}}}_{#1}} }

\def\emptyarrow{\raisebox{.6ex}{
\begin{tikzpicture}
\draw[-open triangle 60](0,0)-- (0.7,0);
\end{tikzpicture}
}}
\def\filledarrow{\raisebox{.6ex}{
\begin{tikzpicture}
\draw[-triangle 60](0,0)--(0.7,0);
\end{tikzpicture}
}}

\newcommand{\LRttrans}{\ensuremath{\stackrel{\ast}{\longleftrightarrow}}}
\newcommand{\ttrans}[1][]{\ensuremath{\stackrel{#1}{\longrightarrow}}}
\newcommand{\sseq}[1][]{\hspace{-1ex}\ensuremath{\stackrel{#1}{\emptyarrow}}\hspace{-1ex}}
\newcommand{\ssync}[1][]{\hspace{-1ex}\ensuremath{\stackrel{#1}{\filledarrow}}\hspace{-1ex}}
\newcommand{\ttranss}[0]{\trans[\ast]}
\newcommand{\sseqq}[0]{\seq[\ast]}
\newcommand{\ssyncc}[0]{\sync[\ast]}
\newcommand{\LRtrans}{\ \LRttrans\ }
\newcommand{\trans}[1][]{\ \ttrans[#1]\ }
\newcommand{\seq}[1][]{\ \sseq[#1]\ }
\newcommand{\sync}[1][]{\ \ssync[#1]\ }
\newcommand{\transs}[0]{\ \ttranss\ }
\newcommand{\seqq}[0]{\ \sseqq\ }
\newcommand{\syncc}[0]{\ \ssyncc\ }

\newcommand{\cparbox}[2]{\parbox{#1}{\begin{center}#2\end{center}}}
\newcommand{\rightbox}[2]{\par\hfill\parbox{#1}{#2}}
\newcommand{\spacebox}[3][t]{\parbox[#1]{#2}{\begin{spacing}{0}#3\end{spacing}}}
\newcommand{\scaleparbox}[3]{\scalebox{#1}{\parbox{#2}{#3}}}
\newcommand{\scaleparboxS}[4]{\scalebox{#1}[#2]{\parbox{#3}{#4}}}

\setlength{\unitlength}{1cm}

\newcommand{\pardessus}[3]
{
\begin{picture}(0,0)
\put(#1,#2){#3}
\end{picture}
}

\newcommand{\pardessusbox}[4]
{
\pardessus{#1}{#2}{\parbox{#3}{#4}}
}

\newcommand{\inserred}{\rotatebox{90}{\ensuremath{\succ}}}
\newcommand{\inserreu}{\rotatebox{90}{\ensuremath{\prec}}}
\newcommand{\inserreD}{\scalebox{3}[1.5]{\rotatebox{90}{\ensuremath{<}}}}
\newcommand{\inserreU}{\scalebox{3}[1.5]{\rotatebox{90}{\ensuremath{>}}}}

\newlength{\udnotelength}

\newcommand{\unote}[4][gray]{\hspace{#2} \cparbox{#3}{\footnotesize\color{#1} #4\par\vspace{-1mm}\scalebox{2}[1.3]{\inserreu}}\par\vspace{-4mm}}

\newcommand{\unoteH}[5][gray]{\pardessusbox{#2}{#3}{#4}{\unote[#1]{0mm}{#4}{#5}}}
\newcommand{\unoteHH}[4][gray]{
  \settowidth{\udnotelength}{#4}%
  \pardessusbox{#2}{#3}{\linewidth}{\hspace{-0.5\udnotelength}\cparbox{\udnotelength}{\footnotesize\color{#1}#4\par\vspace{-1.7mm}\scalebox{1.5}[1.1]{\inserreu}}}
}

\newcommand{\dnoteH}[4][gray]{
  \settowidth{\udnotelength}{#4}%
  \pardessusbox{#2}{#3}{\linewidth}{\hspace{-0.5\udnotelength}\cparbox{\udnotelength}{\footnotesize\color{#1}\scalebox{1.5}[1.1]{\inserred}\par\vspace{-2mm}#4}}
}

\newcommand{\whitedot}[0]{\textcolor{white}{.}}
\newcommand{\hfillS}[0]{\textcolor{white}{.}\hfill}
\newcommand{\hspaceS}[1]{\textcolor{white}{.}\hspace{#1}\textcolor{white}{.}}

\newcommand{\kauses}[0]{\ensuremath{\lhd\hspace{-5pt}\lhd~}}
\newcommand{\kausesnot}[0]{\ensuremath{\causes\hspace{-11pt}/\hspace{13pt}}}
\def\STl{{\sc shortest traj.} length}
\newcommand{\moov}[2][z]{\ensuremath{\nabla#1_{#2}}}

\newcommand{\refacyclic}[0]{Lemma~\ref{short}~\ref{Gacyclic}~}

\newcommand{\urlPR}{http://perso.ens-lyon.fr/mathilde.noual/application/projetlong.pdf}
\newcommand{\urldossier}{http://perso.ens-lyon.fr/mathilde.noual/application/}

\definecolor{newcol}{RGB}{0,75,100}
\colorlet{newcolC}{newcol}
\colorlet{newcolL}{newcol}
\colorlet{newcolU}{newcol}
\newcommand{\mine}[1]{\colorlet{newcolU}{ForestGreen!70!green} #1\colorlet{newcolU}{newcol}}
\newcommand{\mineb}[0]{\colorlet{newcolU}{ForestGreen!70!green}}
\newcommand{\minee}[0]{\colorlet{newcolU}{newcol}}

\newcommand{\neuf}[1]{\textcolor{red}{#1}}



\author{M. Noual}
 \title{Shortest Trajectories and Reversibility\\ in Boolean Automata Networks\\[3mm] {\small
 M. Noual}}
\maketitle







\section{Introduction}

\pgh{Intuition and motivation.}
It seems that often, when a Boolean Automata Network (BAN) $\ntk$ can make a
global change, it can do it rather quickly, \ie{} with few local changes.  More
precisely, when it is possible to reach a certain specific configuration
$y\in\Bn=\set{0,1}^n$ of $\ntk$, starting from an initial configuration
$x\in\Bn$ of $\ntk$, then it seams that the following is often the case. To make
the global change $x\leadsto y$, only a small number of local changes  need to be made, \ie{} only a small number of
automata need to 'move' (change states): something (polynomially)
comparable to the size $n\in\N$ of the network $\ntk$ \ie{} to the number $n$ of
automata in $\ntk$ and  to the total number of different conceivable
automata moves away from $x$.
To check this conjecture and specify the meaning of ``{\em often}'' in this
context, we take interest here in ``{long} trajectories''.
\pgh{{Long} trajectories.} To qualify as {long}, a trajectory
must switch {\em some} automata state values back and forth between $0$ and
$1$. In a trajectory that isn't long, every automaton $i\in
V=\set{1,\ldots,n}$ of the BAN $\ntk$ either doesn't move at all, or only
moves once. The whole length of a trajectory that isn't long (the number of automata
moves it involves) is no greater than the total number $n$ of automata in $\ntk$.

\pgh{Long {shortest} trajectories.} 
We are interested in the case where to get from a configuration $x\in\Bn$ to a
configuration $y\in\Bn$, there is no shorter way than to have some automata
moving back and forth. In other terms, all shortest trajectories from $x$ to $y$
are long. In such cases, we will say that to get from $x$ to $y$ requires ``{reversibility}''.


\pgh{Moves.} Formally, in this Boolean context, a {move} of an
automaton $i\in V$ is the transition of its actual state, the state
$x_i\in\B=\{0,1\}$ that $i$ has in configuration $x=(x_1,x_n, \ldots,
x_n)\in\Bn$, to the only other different state $i$ can take, namely $\neg x_i\in\B$. So
a move of $i\in V$ is either $x_i=0\leadsto \neg x_i=1$, or $x_i=1\leadsto \neg
x_i=0$.
The first (resp. second) kind of move is represented by the value $+1$
(resp. $-1$).  Generally, we write $\nabla x_i=- \nabla \neg x_i=\neg
x_i-x_i\in \SSS=\set{-1,+1}$. $\nabla x_i$ (resp. $-\nabla x_i$) 
represents $i$'s move {\em away} from (resp. {\em towards}) $x$.  If ever automaton $i\in V$ has the capacity to make a move in configuration $x\in\Bn$, then
this move necessarily is the move represented by  $\nabla x_i\in\SSS$. 

\pgh{Signs and Boolean values.} We introduce function
\mbox{$\SB:\SSS\to \B$} defined by 
\mbox{$\SB(s)=\frac{s+1}{2}$} so that
for any configuration $x\in\Bn$, $\SB(-\nabla x_i)=x_i$ equals the state
 of automaton $i$ in configuration $x$. And \mbox{$\SB^{-1}=\BS:
\B\to \SSS$}  so that \mbox{$\BS(x_i)=  2x_i-1=-\nabla x_i$} is $i$'s move
 towards  $x$.

\pgh{BANs (Boolean Automata Networks).} It remains the question: {\em In
$x\in\Bn$, can automaton $i$ make move $\nabla x_i$ or can it not?} Precisely,
this is determined by the definition of the BAN. A BAN ${\cal N}$ is a set of
{\em local transition functions}: $\ntk=\{f_i:\Bn\to \B,\ i\in V\}$, one for
each automaton in the set $V=\iset{1,n}$ of all the BAN's automata.

\pgh{Instability and stability.} In configuration $x$, the automaton
$i\in V$ can make move $\nabla x_i$ if and only if the following holds: $f_i(x)
= \neg x_i$ and equivalently $\nabla x_i=f_i(x)-x_i$. In this case, $i$ is said
to be {\em unstable in configuration $x$}. The set of automata that are unstable
in $x$ (ready to make a move in $x$) is  $U(x)=\{i\in V:\ f_i(x) \neq
x_i\}$.
If automaton $i\in V$ cannot make  move 
$\nabla x_i$ in configuration $x$, \ie{} if $f_i(x)-x_i= 0$, then
$i$ is said to be {\em stable in $x$}. The set of automata that are stable in $x$ is
 $S(x)=\{i\in V:\ x_i=f_i(x)\}$.

\pgh{Signature of punctual influences.} Let $i,j\in V$ be two 
{\em different} automata of the same BAN $\ntk$. We define the {sign} of the influence that
automaton $j$ has on automaton $i$ in configuration $x\in\Bn$ as follows
\textcolor{gray!70}{(where configuration $\bar{x}{j}=(x_1,\ldots,
x_{j-1},\neg x_j, x_{j+1},\ldots )$ is exactly the same as configuration $x$
except for component $j$: $\bar{x}{j}_j=\neg x_j$ and $\forall
k\neq j,\ \bar{x}{j}_k= x_k$)}. \\
$\forall x\in\Bn,\forall i\neq j\in
V,\sign(x,j,i)=(f_i(\bar{x}{j}) -f_i(x))\cdot
\textcolor{gray!70}{\underbrace{\color{black}(\bar{x}{j}_j -
x_j)}_{= \nabla x_j}}
\in \SSS\cup \set{0}$, \ie{} 

$\sign(x,j,i)=\begin{cases}
0 & \text{if }i\in U(x)\cap U(\bar{x}{j}) \\[-1.3mm]
& \pardessus{0}{0}{\text{\textcolor{gray!70}{because then: $f_i(x)=\neg
x_i=\neg \bar{x}{j}_i = f_i(\bar{x}{j})$}}}\\
-\nabla x_i\cdot \nabla x_j & \text{if }i\in U(x)\cap S(\bar{x}{j})  \\[-1.3mm]
& \pardessus{0}{0}{\text{\textcolor{gray!70}{because then: $f_i(x)= \neg x_i$ and 
$f_i(\bar{x}{j})= \bar{x}{j}_i =\neg x_i$}}}\\
\nabla x_i\cdot
\nabla x_j & \text{if } i\in S(x)\cap U(\bar{x}{j})  \\[-1.3mm]
& \pardessus{0}{0}{\text{\textcolor{gray!70}{because then: $f_i(x)=x_i$ and 
$f_i(\bar{x}{j})= \neg \bar{x}{j}_i =\neg x_i$}}}\\
0 & \text{if }i\in S(x)\cap S(\bar{x}{j})   \\[-1.3mm]
& \pardessus{0}{0}{\text{\textcolor{gray!70}{because then: $f_i(x)=
x_i= \bar{x}{j}_i = f_i(\bar{x}{j})$}}}
\end{cases}$
\medskip

When $j=i$, we define the sign of the influence automaton $i$ has on itself as
follows.\\[2mm]
$\forall x\in\Bn,\ \forall i\in V,$
$\sign(x,i,i)=(f_i(\bar{x}{i}) -f_i(x))\cdot
\textcolor{gray!70}{\underbrace{\color{black}(\bar{x}{i}_i - x_i)}_{\nabla x_i}}$,~ \ie:\\[2mm]
$\sign(x,i,i)=\begin{cases}
-\nabla x_i\cdot \nabla x_i= -1  & \text{if }i\in U(x)\cap U(\bar{x}{j})   \\[-1.3mm]
& \pardessus{0}{0}{\text{\textcolor{gray!70}{because then: $f_i(x)=\neg
x_i\neq x_i = f_i(\bar{x}{i})$}}}\\
0 & \text{if }i\in U(x)\cap S(\bar{x}{j})   \\[-1.3mm]
& \pardessus{0}{0}{\text{\textcolor{gray!70}{because then: $f_i(x)= \neg x_i=f_i(\bar{x}{i})$}}}\\
0 & \text{if } i\in S(x)\cap U(\bar{x}{j})   \\[-1.3mm]
& \pardessus{0}{0}{\text{\textcolor{gray!70}{because then:
$f_i(x)=x_i= \neg \bar{x}{i}_i= f_i(\bar{x}{i})$}}}\\
\nabla x_i\cdot \nabla x_i=+1 & \text{if }i\in S(x)\cap S(\bar{x}{j})   \\[-1.3mm]
& \pardessus{0}{0}{\text{\textcolor{gray!70}{because then: $f_i(x)=
x_i\neq   f_i(\bar{x}{i})$}}}
\end{cases}$

\pgh{Monotone  functions and Monotone BANs.}
A Boolean function $f:\Bn\to \B$ is said to be monotone when the following
holds. In the conjunctive normal expression of $f(x)$, a literal $x_j$ either
only appears unnegated, or it only appears negated. In a BAN $\ntk$, the
monotony of $f_i$ is equivalent to: $\forall x,y\in\Bn, \sign(x,i,j)\neq
0 \wedge \sign(y,i,j)\neq 0 \implies \sign(x,i,j)=\sign(y,i,j)$. In that case,
we write $\sign(i,j)=\sign(x,i,j)$ $ \forall x\in\Bn$ s.t. $\sign(x,i,j)\neq 0 $.
A monotone BAN is one in which all local transition functions $f_i$ are
monotone. 

\pgh{Interaction Graph.} The interaction graph of a BAN $\ntk$ is  the
digraph $G=(V,A)$ where $A=\set{(j,i)\in V\times V\,:\, \exists
 x\in \Bn,\ \sign(x,j,i)\neq 0}$. We say that a BAN $\ntk$ is (strongly)
 connected if its interaction graph $G$ is (strongly)
 connected.

\pgh{Path signs.} In a monotone BAN, the sign of a path of $G$ is the product of the signs of
 the arcs it is comprised of. 

\pgh{Monotone and contradictory paths.} In  $G$, if there 
are no paths from automaton $j$ to automaton $i$, we let $\sign^\ast(j,i)=
0$. Otherwise, if all paths from automaton $j$ to automaton $i$ have the same sign
$s$, then we let $\sign^\ast(j,i)= s$.  And if there exists a negative path as
well as a positive path from $j$ to $i$, we say that these two paths are
{contradictory paths}. In particular, the path covering a negative cycle once,
and the path covering it twice are contradictory paths.

\pgh{Nice Networks and totally positive ones.} A {nice BAN} is a BAN for which $\sign^{\ast}(i,j)$
 is defined $\forall i,j \in V$. In other terms, a nice BAN is a monotone BAN  without
 contradictory paths, and in particular without negative cycles. 
A totally positive BAN is a monotone (and nice) BAN in which $\forall i,j \in
V$: $\sign(i,j)=+1$.

\pgh{An important remark on the meaning of automata state values.} 
 In BANs, the values
$0$ and $1$ of automata states are often assumed to represent two opposite state
values, the same for all automata, so that when state $x_i$ of automaton $i$
equals $0$, and state $x_j$ of automaton $j$ also equals $0$, then the equality
$x_i=x_j$ is taken to mean that the same thing is happening to $i$ and $j$.\linebreak
Often, $x_i=x_j=0$ is taken to mean that $i$ and $j$ are both ``inactive'', as
opposed to ``active'' as they would be if they were in state $1$. In reality, in
a BAN, $0$ and $1$ are just labels. $x_i=0$ could just as well mean that $i$ is
``open'' rather than ``closed'', while for the automaton $j$ next door, $x_j=0$
would  mean that $j$ is ``blue'' rather than ``black''.
Thus, $x_i=x_j=0$ only is meaningful in the sense of $x_i=0$ and $x_j=0$,
not in the sense of $x_i=x_j$.\linebreak
%
There are no other relations between $x_i=0$ and $x_j=0$ than the
ones ensuing from the definitions of the BAN's local transitions functions. 
And  since the values $0$ and $1$ of automata states are just {\em names} of automata state
values, this means that we can  exchange state $0$ of automaton $i$ with state
$1$, for instance. Replacing \litem every $f_j(x)$ by $f_j(\bar{x}{i})$ and
then \litem\endlitem $f_i(x)$ by
$\neg f_i(x) $ yields different  a formalisation of the exact same network. In this new
formalisation, all arcs $(j,i)\in A$ and $(i,j)\in A$, $j\in V$  have changed signs.

\pgh{}\label{nicetopositive}
It was proven in \cite{Melliti2013} that any  nice strongly connected BAN can be
re-formulated as a  totally positive BAN.

\pgh{Neighbourhoods.}
The {in- (resp. out-) neighbourhood} of an automaton
$i\in V$  is the set $V_{\to i}=\set{j\in V:\ (j,i)\in A}$ (resp. the set $V_{i\to}=\set{j\in V:\ (i,j)\in A}$).

\pgh{Neighbour Inputs and Straight functions.}\label{straight}
In a monotone BAN, $\forall j\in V_{\to i}$, $\sign(j,i)$ is defined. 
We write  $x_{j\to
i}=\SB(\,\sign(j,i)\cdot \BS(x_j) \,)=
\SB(\,-\sign(j,i)\cdot \nabla
x_j \,)\in\B$ to denote the Boolean value incoming automaton $j$ that automaton
$i$ computes with in configuration $x$, \ie{} the {\em input} that $i$ receives in 
$x$ from  $j$:

$$
x_{j\to i}=\begin{cases} 
f_i(x)= \neg x_i & \text{if } i\in U(x) \text{ and } \sign(x,j,i)=\sign(j,i)\\
f_i(x)= x_i & \text{if } i\in S(x) \text{ and } \sign(x,j,i)=\sign(j,i)\\
\neg f_i(x) = x_i & \text{if } i\in U(x) \text{ and } \sign(x,j,i)=0\\
\neg f_i(x) = \neg x_i & \text{if } i\in S(x) \text{ and } \sign(x,j,i)=0\\
%
\end{cases}
$$
When $j\notin V_{\to i}$, we let $x_{j\to i}=x_j$.  With 
$x_{j\to i}$ thus defined for all $x$,  we define the
``straight'' local transition function $g_i$ associated to any local transition function $f_i$ of the BAN:
$$
\forall x\in \Bn, g_i(x_{1\to i},\ldots x_{n\to i})=f_i(x).
$$
The interest in $g_i$ is that its conjunctive normal form contains no
negation. The negations of $f_i$ are all already taken into account by changing
$x_j$ for $x_{j\to i}$. And thus we see that $f_i(x)=b\in\B\implies \exists j\in
V_{\to i}:\, x_{j\to i}=b$.


\pgh{(Asynchronous) Trajectories.} 
A {trajectory} from $x=(x_1,\ldots,x_n)\in\Bn$ to $y=(y_1,\ldots,y_n)\in\Bn$
is a sequel of configurations $(x(t))_{t\leq T}$ such that $x(0)=x$, $x(T)=y$,
and $\forall t<T$, configurations $x(t)$ and $x(t+1)$ are related as follows:
$x(t+1)=\bar{x(t)}{i}$ for some automaton $i\in U(x(t))=U(t)$ that is unstable
at time step $t$. We denote this automaton by $i=\nu(t)$. Thus: $
\forall t\in\isetl{0,T},\ x(t+1)=\bar{x(t)}{\nu(t)}$.

\pgh{Recurrent configurations.} Let $x\in\Bn$ be a configuration of the
BAN $\ntk$. Consider the set of all configurations $y\in\Bn$ such that there is
a trajectory from $x$ to $y$. If all configurations $y$ in this set are such
that there also is a trajectory back from $y$ to $x$, then we call $x$ a
{recurrent configuration} ($y$ necessarily is one too). Stable configurations
$x\in \Bn$ s.t. $S(x)=V$ are special kinds of recurrent configurations.

\pgh{}From now on, we consider an arbitrary trajectory $\traj=(x(t))_{t\in \alltimes}$ from
$x=x(0)$ to $y=x(T)$, where $\alltimes=\iset{0,T}$ is the set of all times steps
of the trajectory and $T\in\N$ is the last of them.


\section{Causality (version 0.1)}
\pgh{Causality.} 
Consider time step $t$ of $\traj$ at which move $\nabla x(t)_i$ is made by
automaton $i=\nu(t)\in U(t)=U(x(t))$. We want  to pinpoint what ``{caused}'' or 
''{unlocked}'' the possibility of making this move. There are two
cases:\smallskip

\begin{enumerate}[leftmargin=0cm,labelsep=0.3cm]
\item Move $\nabla x_i(t)$ was  possible ever  since the beginning of $\traj$:
$\forall t' <t,\ i\in U(t')$ and $\nu(t')\neq i$. 
%
%
In this case, $t$ is said to be a {\em root step}, and move $\nabla x(t)_{i}$ a
{\em root move}. A root move  happening at a root step has no cause.

\item There exists a step $t'<t$ at which  move $\nabla x(t)_{i}$
wasn't yet possible: $i\in S(t')$ was stable and in state
$x(t')_i=x(t)_i$. 
We denote by: $$\tau(t)=max\{t'<t\,:\, \forall s\in \isetr{t',t},\, i\in
U(s)\cap S(t')\text{ and } \nu(s)\neq i\}$$ the most recent time step at which
$i$ was stable before $t$. 
At time step $\tau(t)$ a move $\nabla x(\tau(t))_j$ was made by a certain
 automaton $j=\nu(\tau(t))$. Right after that, $i$
 became unstable. And it remained unstable and unmoved until  the time step $t$ at
 which it made move  $\nabla
 x(t)_{i}=\nabla x(\tau(t))_{i}= \nabla
 x(s)_{i}, \,\forall s\in \iset{\tau(t),t}$.
%
Automaton $j$'s move $\nabla x(\tau(t))_{j}$ at time $\tau(t)$ is said to be the
 {cause} of automaton $i$'s move $\nabla x(t)_{i}$  at time $t$.
\end{enumerate}

\begin{lemma}
\label{signs tau}{\color{white}.}

{\color{black!70}{Let ${\cal T}=(x(t))_{t\in\alltimes }$ be a trajectory of a BAN
$\ntk=\{f_i:\Bn\to \B,\ i\in V\}$.}\linebreak Let $t_1<t_2<T$ be two time steps
of ${\cal T}$. Let $j=\nu(t_1)$ and $i=\nu(t_2)$.}\linebreak\vspace{-3.3mm}

 If $t_1=\tau(t_2)$, then
$\sign(x(t_1),j,i)$ 
$= \nabla x(t_1)_{j} \cdot \nabla x(t_1)_{i}= \nabla x(t_1)_{j}\cdot \nabla x(t_2)_{i}$.
\end{lemma}

\begin{myproof}
 $\bar{x(t_1)}{j}=x(t_1+1)$ and $i\in S(x(t_1))\cap  U(x(t_1+1))$.
\end{myproof}

\pgh{Causality branches.} $\forall t\in\alltimes $, we let $\tau^1(t)=\tau(t)$, and $\forall
q\in\N,\ \tau^{q+1}(t)=\tau(\tau^q(t))$ if it exists. It doesn't if $\tau^q(t)$
is a root step. And if $p=max\{q\in\N\text{ such that } \tau^{q}(t)\text{ exists}\}$, then we denote by $\tau^{\ast}(t)=\tau^{p}(t)$ this root step. 

A {$\tau$-branch} of   trajectory ${\cal T}$ is a sequel of $q\in\N$ time steps
$(\tau^{q-p}(t))_{p\leq q}$ ending with time step $t<T$.
\medskip

The following lemma relates path signs in $G$ to moves made  along $\tau$-branches.

\begin{lemma}
\label{signs paths tau}
{\color{black!70}Same conditions as Lemma~\ref{signs tau}.}\\  If $\exists
q\in\N:\, t_1=\tau^q( t_2)$, then there is a path of length $p$ in $G$
from $j$ to $i$ \linebreak
 {\color{black!80}(which  has sign
$\nabla x(t_1)_{j}\cdot \nabla x(t_2)_{i}$ if $\ntk$ is monotone)}.
If $j=i$, this path is a cycle and all
automata that move on the same branch $(\tau^p(t_2))_{p\leq q}$ between $t_1$
and $t_2$ are strongly connected in $G$.
\end{lemma}

\begin{myproof}By induction on $q$. If $q=1$, then it follows from Lemma~\ref{signs
 tau}. If $q>1$, then there exists $t\in \iseto{t_1,t_2}\,:\,
 t_1=\tau(t)$ and $t=\tau^{q-1}(t_2)$.  By the induction hypothesis, there exists
 a path of length $q-1$ {\color{black!70}(which has  sign
 $\nabla x(t)_{k} \nabla x(t_2)_{i}$ if $\ntk$ is monotone)} from $k=\nu(t)$ to
 $i=\nu(t_2)$ in $G$. There also exists  an arc $(j,k)\in A$ {\color{black!70} (of sign
 $\nabla x(t_1)_{j} \nabla x(t)_{k}$)}.  The concatenation of this path and
 this arc defines a path from $j$ to $i$ of length $q+1$
 {\color{black!70}(and sign $\nabla x(t_2)_{i}\nabla x(t_1)_{j}$)}.
\end{myproof}

The next result is a direct consequence of Lemma~\ref{signs paths tau}.

\begin{lemma}
\label{branches tau}{\color{white}.}

{\color{black!70}Let $\traj=(x(t))_{t\in\alltimes}$ be a trajectory of a BAN
$\ntk=\{f_i:\Bn\to \B,\ i\in V\}$. }\linebreak
 If there exists an automaton moving up and down along a $\tau$-branch of
$\traj$,  then
$\ntk$ is not nice: either it is non-monotone, or there is a negative cycle in
$G$ involving $i$.
\end{lemma}

\begin{myproof}The existence of an automaton moving up and down along a $\tau$-branch of
$\traj$ is equivalent to the following.  $\exists p\in\N,t_1,t_2\in \alltimes \,:\, t_2=\tau^p(t_1)$
and $\nu(t_1)=\nu(t_2)=i$ and $\nabla x(t_1)_{i}=-\nabla x(t_2)_{i}$.
\end{myproof}

\pgh{Causality trees.} A {$\tau$-tree} is a set of
{$\tau$-branches} sharing  their smallest time step. The smallest time step of a
maximal {$\tau$-tree}  is a root step. There are
as many maximal $\tau$-trees as there are root moves, so no more than
$|U(0)|\leq n$. \medskip

The following result is a second consequence of Lemma~\ref{signs paths tau}.

\begin{lemma}
\label{trees tau}
{\color{black!70}Same conditions as Lemma~\ref{branches tau}.}\\ If there exists
 an automaton $i$ moving up and down on the same $\tau$-tree of ${\cal T}$, then
 $\ntk$ is not nice: either it is non-monotone or there are contradictory paths
 in $G$ (from  $\nu(t_0)$ to $i$ where $t_0$ is the smallest time-step
 of the $\tau$-tree).
\end{lemma}

\begin{myproof}The existence of an automaton $i$ moving up and down on  a $\tau$-tree of
$\traj$ is equivalent to the following.  $\exists p,q\in\N,
 t_0,t_1,t_2\in \alltimes \,:\, t_0=\tau^{p}(t_1)=\tau^{q}(t_2)$ and
 $\nu(t_1)=\nu(t_2)=i$ and $\nabla x(t_1)_{i}=-\nabla x(t_2)_{i}$.
\end{myproof}

\begin{lemma}
{\color{black!70}Same conditions as Lemma~\ref{branches tau}.}\\
If $\ntk$ is a nice BAN and if  an automaton makes twice  the
 same move on a $\tau$-branch of ${\cal T}$,  then there are  at least
two $\tau$-trees in ${\cal T}$ and a positive cycle in $G$. 
\end{lemma}

\begin{myproof}
If automaton $i$ goes up twice on the same branch {\color{gray}(making move
$\nabla_i$)}, then, by Lemma~\ref{signs paths tau}, it belongs to a cycle  which
must be positive because of $\ntk$'s niceness. And $i$ must go back down once
{\color{gray}(making move $-\nabla_i$)} between every time it goes up. The
niceness of $\ntk$ and Lemma~\ref{branches tau} imply that this must happen on
another tree.
\end{myproof}

\pgh{}If $\ntk$ is totally positive, then there are only two types of trees: {\it (i)}
trees comprised of $\nabla x(t)_i=1$ moves, \ie{} trees on which automata in state
$0$ move to state $1$ and {\it (ii)} trees comprised of $\nabla x(t)_i=0$
moves, \ie{} trees on which automata in state $1$ move to state $0$.

\pgh{$G_{\tau}$ and the anti-graph of the $\tau$ function.}  By definition of $\tau$,
$\forall t\in \alltimes $ there is at most one $t'\in \alltimes $ satisfying
$t'=\tau(t)$. We can therefore consider the graph of function $\tau^{-1}$ a.k.a.
{anti}-graph of function $\tau$. This is the digraph with node set $\alltimes $
and arc set $\{(\tau(t),t),\ t\in \alltimes \}$). 
Let us call {\em strongly acyclic} a digraph whose undirected version is acyclic.
The anti-graph of $\tau$ is strongly acyclic.
A $\tau$-tree is a connected  component of the anti-graph of $\tau$.
A $\tau$-branch  is a  linear sub-graph of the anti-graph of $\tau$.
We can also naturally define $G_{\tau}=(V,A_{\tau})$ where
$A_{\tau}=\{\,(\nu(\tau(t)),\nu(t)),\ t\in \alltimes \}$. By Lemma \ref{signs tau},
$G_{\tau}$ is  a subgraph of the interaction graph $G$. 
\section{Hamiltonian shortest trajectories}

\pgh{Hamiltonian shortest trajectories.} A {Hamiltonian} shortest 
trajectory of a BAN $\ntk$, is a {shortest} trajectory ${\cal T}$ of $\ntk$
that goes from one of its configurations $x$ to another $y$ by going through
all other configurations of $\ntk$. A {Hamiltonian} shortest trajectory of a BAN
of size $n$ has length $T=2^n$.

\begin{proposition}
\label{Hamiltonian}
If a BAN has a {Hamiltonian} shortest trajectory then it is not a nice BAN.
\end{proposition}

\begin{myproof}Let ${\cal T}=(x(t))_{t\in\alltimes }$ be a {Hamiltonian} shortest trajectory of
$\ntk$. \linebreak Let $i=\nu(0)$ be the first automaton to make a move along
$\traj$. If $\exists j\in U(0)\setminus\{i\}$, then this $j$ can make a move in
configuration $x(0)$ and $x(0)\longrightarrow \bar{x(0)}{j}$ is a transition
that $\ntk$ can make, which shortcuts ${\cal T}$. Indeed, since ${\cal T}$ is
Hamiltonian, $\exists t\in\alltimes $ such that $x(t)=\bar{x(0)}{j}$. And thus the
transition $x(0)\longrightarrow \bar{x(0)}{j}=x(t)$ takes $\ntk$ straight from
$x(0)$ to $ x(t)$ without going through configuration
$x(1)= \bar{x(0)}{i} \neq \bar{x(0)}{j}=x(t)$.  But ${\cal T}$ cannot be
shortcut if ${\cal T}$ is a shortest trajectory. Thus $U(0)=\{i\}$. ${\cal T}$
has a unique root, and thereby a unique $\tau$-tree. Proposition~\ref{Hamiltonian}
results from this, from Lemma~\ref{trees tau}, and from the fact that along a {Hamiltonian} trajectory (which has
length $2^n>n$) there necessarily are automata moving up and down.
\end{myproof}

\pgh{} This settles the case of the {longest} type of long shortest
trajectories:\linebreak they only exist in non-nice BANs.
Still, nice BANs can have long (non-Hamiltonian) shortest trajectories as the
following example shows.\medskip

\fbox{\parbox{\linewidth}{
\begin{example}\normalfont
\label{totally positive example}
{\small
$\ntk=\{f_i,\ i\in V=\{1,\ldots, 5\}\}$ where:\medskip

\begin{tabular}{p{7.7cm}p{4cm}}
\hspace{2cm}$
\begin{cases}
f_1 &=~ x_4 \wedge x_5\nonumber\\
f_2 &=~ x_1 \vee x_2\nonumber\\
f_3 &=~ (x_1 \vee x_2) \wedge x_4\nonumber\\
f_4 &=~ x_3\nonumber\\
f_5 &=~ x_1\vee x_3 \vee x_4\nonumber
\end{cases}
$\\[-3.5cm]&
\hfill\scalebox{0.4}{\begin{picture}(0,0)%
\includegraphics{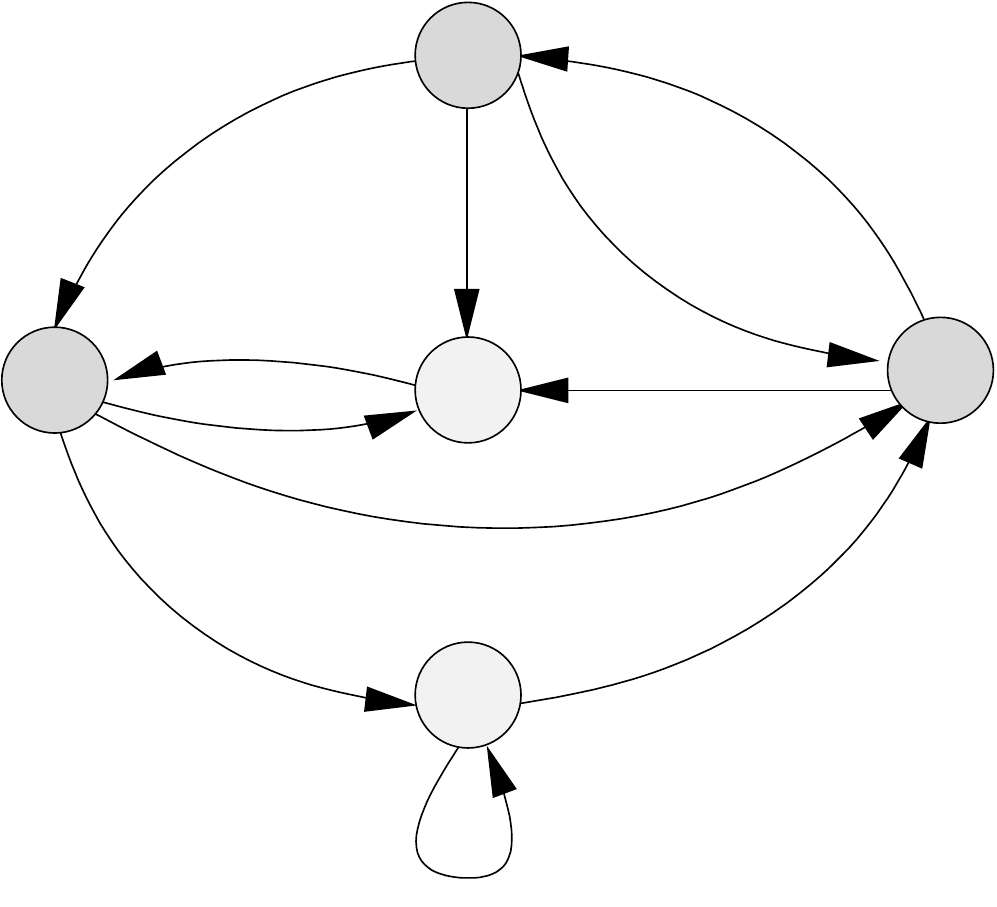}%
\end{picture}%
\setlength{\unitlength}{4144sp}%
\begingroup\makeatletter\ifx\SetFigFont\undefined%
\gdef\SetFigFont#1#2#3#4#5{%
  \reset@font\fontsize{#1}{#2pt}%
  \fontfamily{#3}\fontseries{#4}\fontshape{#5}%
  \selectfont}%
\fi\endgroup%
\begin{picture}(4550,4140)(3897,-5075)
\put(5986,-5011){\makebox(0,0)[lb]{\smash{{\SetFigFont{12}{14.4}{\rmdefault}{\mddefault}{\updefault}{\color[rgb]{0,0,0}\colorbox{white}{$+$}}%
}}}}
\put(7156,-3886){\makebox(0,0)[lb]{\smash{{\SetFigFont{12}{14.4}{\rmdefault}{\mddefault}{\updefault}{\color[rgb]{0,0,0}\colorbox{white}{$+$}}%
}}}}
\put(7516,-1726){\makebox(0,0)[lb]{\smash{{\SetFigFont{12}{14.4}{\rmdefault}{\mddefault}{\updefault}{\color[rgb]{0,0,0}\colorbox{white}{$+$}}%
}}}}
\put(5941,-1861){\makebox(0,0)[lb]{\smash{{\SetFigFont{12}{14.4}{\rmdefault}{\mddefault}{\updefault}{\color[rgb]{0,0,0}\colorbox{white}{$+$}}%
}}}}
\put(6661,-2131){\makebox(0,0)[lb]{\smash{{\SetFigFont{12}{14.4}{\rmdefault}{\mddefault}{\updefault}{\color[rgb]{0,0,0}\colorbox{white}{$+$}}%
}}}}
\put(6976,-2761){\makebox(0,0)[lb]{\smash{{\SetFigFont{12}{14.4}{\rmdefault}{\mddefault}{\updefault}{\color[rgb]{0,0,0}\colorbox{white}{$+$}}%
}}}}
\put(4726,-1636){\makebox(0,0)[lb]{\smash{{\SetFigFont{12}{14.4}{\rmdefault}{\mddefault}{\updefault}{\color[rgb]{0,0,0}\colorbox{white}{$+$}}%
}}}}
\put(5131,-2626){\makebox(0,0)[lb]{\smash{{\SetFigFont{12}{14.4}{\rmdefault}{\mddefault}{\updefault}{\color[rgb]{0,0,0}\colorbox{white}{$+$}}%
}}}}
\put(4591,-3751){\makebox(0,0)[lb]{\smash{{\SetFigFont{12}{14.4}{\rmdefault}{\mddefault}{\updefault}{\color[rgb]{0,0,0}\colorbox{white}{$+$}}%
}}}}
\put(5986,-3391){\makebox(0,0)[lb]{\smash{{\SetFigFont{12}{14.4}{\rmdefault}{\mddefault}{\updefault}{\color[rgb]{0,0,0}\colorbox{white}{$+$}}%
}}}}
\put(4951,-2941){\makebox(0,0)[lb]{\smash{{\SetFigFont{12}{14.4}{\rmdefault}{\mddefault}{\updefault}{\color[rgb]{0,0,0}\colorbox{white}{$+$}}%
}}}}
\put(5941,-2806){\makebox(0,0)[lb]{\smash{{\SetFigFont{20}{24.0}{\rmdefault}{\mddefault}{\updefault}{\color[rgb]{0,0,0}$5$}%
}}}}
\put(5941,-4201){\makebox(0,0)[lb]{\smash{{\SetFigFont{20}{24.0}{\rmdefault}{\mddefault}{\updefault}{\color[rgb]{0,0,0}$2$}%
}}}}
\put(4051,-2761){\makebox(0,0)[lb]{\smash{{\SetFigFont{20}{24.0}{\rmdefault}{\mddefault}{\updefault}{\color[rgb]{0,0,0}$1$}%
}}}}
\put(5941,-1276){\makebox(0,0)[lb]{\smash{{\SetFigFont{20}{24.0}{\rmdefault}{\mddefault}{\updefault}{\color[rgb]{0,0,0}$4$}%
}}}}
\put(8101,-2716){\makebox(0,0)[lb]{\smash{{\SetFigFont{20}{24.0}{\rmdefault}{\mddefault}{\updefault}{\color[rgb]{0,0,0}$3$}%
}}}}
\end{picture}%
}
\end{tabular}

Let  $x=(x_1,\ldots,x_5)=(1,0,1,1,0)$  and $y=(0,1,0,0,0)$.
To get from configuration $x$ to configuration $y$, $\ntk$ has to move automaton
$1$ up twice and down once. There is no other way:

$ x=x(0)=(1,0,1,1,0) \stackrel{{\color{gray!60}+\nabla x_{1}}}{\longrightarrow}
x(1)=(0,0,1,1,0)\stackrel{{\color{gray!60}+\nabla x_{3}}}{\longrightarrow}x(2)=(0,0,0,1,0)
\stackrel{{\color{gray!60}+\nabla x_{5}}}{\longrightarrow}\ldots
$ $
x(3)=(0,0,0,1,1)\stackrel{{\color{gray!60}-\nabla x_{1}}}{\longrightarrow}x(4)=(1,0,0,1,1)
\stackrel{{\color{gray!60}+\nabla x_{2}}}{\longrightarrow}x(5)=(1,1,0,1,1)\stackrel{{\color{gray!60}+\nabla x_{4}}}{\longrightarrow}
$ $
\ldots
x(6)=(1,1,0,0,1)\stackrel{{\color{gray!60}+\nabla x_{1}}}{\longrightarrow}x(7)=(0,1,0,0,1)
\stackrel{{\color{gray!60}-\nabla x_{5}}}{\longrightarrow}x(8)=(0,1,0,0,0)=y
$\medskip

Remarkably, $\ntk$ is a {totally positive} BAN, and
$y$ is a stable configuration ($S(y)=V$).
}
\end{example}}}

\section{Causality (versions 0.2 and 0.3)}

Assume move $\nabla x(t_1)_{j}$ is made by $j$ at time $t_1=\tau(t_2)$, thereby
unlocking the possibility of move 
$\nabla x(t_2)_{i}$ to be made by $i$ at time $t_1+1$. Assume move $\nabla x(t_1)_{j}$ is
{\em undone} before move $\nabla x(t_2)_{i}$ is done. In other terms, assume move $-\nabla x(t_1)_{j}$
is done at some time $t_3\in \iseto{t_1,t_2}$. Then, for the interval of time
$\isetr{t_3,t_2}$, move $\nabla x(t_2)_{i}$ remains possible despite $j$ no
longer being in the state that unlocked the possibility of this move.
The notion of $\tau$-causality is not such a good notion of causality because of
this. It allows for a caused effect to remain caused even when its cause is no
longer effective.
In particular, with $\tau$-causality, a move causing nothing can be made on a
{\em shortest} trajectory, and later undone. It isn't clear why such a move
would ever be made in the first place. $\tau$-causality does not help understand
why such a move couldn't be skipped to make the trajectory shorter.
Let us call {\em target move} any move $\nabla x_i$ where $i\in
HD(x,y)=\{i\in V: \, x_i\neq y_i\}$. And let us call {\em target-move-causing
move} a move that leads to a target move in a causality
branch.
We would like a notion of causality that allows us to say that shortest
trajectories make no other moves that  {target}
moves and {target-move-causing} moves.  Then, we could say that there are as
many causality branches as there are causality {\em leaves} (branch endpoints),
and as many of those as there are target moves, \ie{} $|HD(x,y)|\leq n$. \smallskip

A second notion of causality, $\kappa$-causality, can be proposed both as an
alternative and a complementary to $\tau$-causality.  In particular, this new
version pinpoints the reason why, although $\tau$-causing nothing, a move may
still be indispensable because it $\kappa$-causes something that is
indispensable.

\pgh{Causality 0.2.}
Let $t\in\alltimes $ and $i=\nu(t)\in V$. $\nabla x(t_1)_{i}$ is the move made at
time step $t_1$ by $i$. $\kappa(t)$ is the set of time steps
at which are made the moves causing (favouring) move $\nabla x(t_1)_{i}$ in the following sense:
$$
t\in \kappa(t_1)\iff \forall s\in\iseto{t,t_1}, \ \nu(s)\neq
j \text{ and } i\in U(t_1)\cap
S(\bar{x(t_1)}{j}).
$$

{\color{black!60}\small
\pardessusbox{2.3}{0.5}{3cm}{$$\underbrace{\hspaceS{1.2cm}}_{\parbox{8cm}{\footnotesize $j$ doesn't move again between
$t$  and $t_1$.}}$$}

\pardessusbox{5}{0.3}{6.5cm}{$$\underbrace{\hspaceS{2.9cm}}_{\parbox{6.3cm}{\footnotesize
The move made at $t_1$ couldn't have been made if the move made at $t$ hadn't
 been made before, or if it had been made but cancelled.}\hspaceS{2.9cm}}$$}}
\vspace{10mm}

Notably, all lemmas above given in terms of $\tau$-causality translate
straightforwardly in terms of
$\kappa$-causality because Lemma~\ref{signs tau} does.\medskip

With this definition, any move of a trajectory ${\cal T}$ that $\kappa$-causes
nothing can be skipped.\medskip

The drawback of this definition is that a move can have several immediate
causes, so it does not yield acyclic causality trees.

\pgh{Causality 0.3.} A third version of causality considers that 
anything is a cause if it is  a cause either by the first version of causality or by
the second version.
\medskip


In the next section, we concentrate on monotone BANs and take a different point of view on
trajectories.

\clearpage
\section{Potentiality}

\pgh{Source Automata.} In the sequel, we are going to assume that $\forall i\in V:\, \deg^-(i)>0$. The
reason is the following. For the sake of keeping notations and developments
simple, we want to avoid having a source automaton $i$ whose local transition
function is constant. Unlike with automata that
have in-neighbours, {the potential} for a source automaton  $i$ to change states is
not carried by any automaton of $V$. Nor is it transmitted through any arc
of $A$.
Let $i\in V$ be an arbitrary source automaton and let $f_i:x\in \Bn\mapsto b$ be
its local transition function. Automaton $i$ can change states only when its
current state  is  $\neg b$. 
So on an arbitrary trajectory ${\cal T}=(x(t))_{t\in\alltimes}$, automaton $i$ moves
if and only if $x(0)_i=\neg b$. If that is the case, $i$ moves at most once. And
when it does, it doesn't inherit its new state from any automaton of $V$. Its
local transition function is the sole responsible actor of the change. Here, we
disregard BANs where such automata exist. Importantly, all targeted
results will apply nonetheless to these BANs  because they are equivalent to BANs in
which each source automaton $i$ is replaced and represented by two automata
$i_1$ and $i_2$ where $f_{i_1}(x)=f_{i_2}(x)=x_{i_1}$ and where initially in
$\traj$, $i_1$ is in state $x(0)_{i_1}=f_i(x)=b$ and $i_2$ is in state
$x(0)_{i_2}=x(0)_i$.
\smallskip

In BANS without real source automata, we call ({\em positive}) {\em source automaton} any
automaton $i\in V:\, V_{\to i}=\set{i} \wedge \sign(i,i)=+1$.  
And we call source loop the cycle defined by the single positive arc $(i,i)\in
A$. Source automata never change states.\smallskip

We let $\V\subset V$ be the set of non-source automata. Those are the only ones
that possibly move along $\traj$: 
$\V\supset  \bigcup_{t\in\alltimes } U(t) \supset \set{\nu(t): t\in\alltimes }$.

\pgh{Potentials.} We are now going to consider couples
$\pot{t,i}\in \potentials=\alltimes \times V$.\linebreak Couple $\pi=\pot{t,i}$ is called {the
 potential carried by automaton $i$ at time $t$}. \linebreak $\pi$ is
 taken to represent {the fact that automaton $i$ is in state $x(t)_i\in \B$
 at time $t$}.
And the idea of this section is to trace back time in order to find and relate
all anterior potentials $\pot{t',j},\, t'<t$ that have participated in
$\pot{t,i}$.

\pgh{Original potential.} Original, or initial potentials are potentials of the form
$\pot{0,i}$ for some $i\in V$. 

\pgh{Equality among potentials.}\label{equality} We define a binary reflexive and transitive  relation $=$ on
$\potentials$ to signify that \litem as long as an automaton $i$ is not moved,
the potential it carries is the same, and \litem\endlitem two different automata
cannot carry the same potential, even at different times:
$$\pot{t,i}=\pot{t',j}\iff i=j \text{ and
} \forall s\in \isetl{t,t'}:\ i\neq \nu(s).$$

\pgh{Transmission and inheritance.}\label{transmission}
We define the {another binary relation}  on
$\potentials$ denoted $\inherits$.
Let $\pi_1=\pot{t_1,j}$ and $\pi_2= \pot{t_2,j}$ be two potentials.
$\pi_2\inherits \pi_2$ means that $\pi_2$ is inherited from $\pi_1$ and 
$j$ transmitted his potential to $i$ in the following sense:
$\forall t\in \alltimes ,\ \forall j\in V, \forall i=\nu(t)\in
V,$\vspace{-2mm}
\begin{eqnarray*}  
 \pot{t,j}\inherits\pot{t+1,i} &\iff &\sign(j,i) =
 {\color{black!60}\underbrace{\color{black}-\nabla
 x(t)_{j}}}~\cdot~
 {\color{black!60}\underbrace{\color{black}\nabla x(t)_{i}}}=\nabla
 x(t)_{j}\cdot\nabla x(t+1)_{i}
\\[10mm]
&\iff& x(t)_{j\to i}=\SB(\,-\sign(j,i) \nabla
x(t)_j \,)=\SB(-\nabla x(t+1)_{i})\\
&\iff& {\color{black!60}\underbrace{\color{black}x(t)_{j\to i}}}~=~{\color{black!60}\underbrace{\color{black}x(t+1)_{i}}}.
\end{eqnarray*} 

{\color{black!60}\small
\pardessusbox{5.5}{2.3}{2cm}{\footnotesize
this $j$-move\\[-1mm] towards $x(t)$\\[-1mm] favours \ldots}
\pardessusbox{7}{2.4}{2.5cm}{\footnotesize \ldots this $i$-move\\[-1mm] away from $x(t)$.}
}

{\color{black!60}\small
\pardessusbox{3.8}{0.4}{2.5cm}{\footnotesize
 $j$ being in\\[-1mm] this state\\[-1mm] favours\ldots}
\pardessusbox{5.3}{0.5}{2.5cm}{\footnotesize \ldots $i$ being \\[-1mm] in this one.}
}\vspace{2mm}

\pgh{}\label{straightagain}By the remark made at the end of \ref{straight}, we
have the third implication below:
\begin{align*}
i=\nu(t) &\implies i\in U(t)\\
&\implies f_i(x(t))=\neg x_i(t)=x_i(t+1)\\
&\implies \exists j\in V_{\to i}:\, x_{j\to i}(t)=x_i(t+1)\iff \exists j\in V_{\to i}:\,\pot{t,j}\inherits\pot{t+1,i}.
\end{align*}
Thus, no automaton can change states without inheriting.

\pgh{}
To define a relation of inheritance,   we could require much stronger
conditions than those implied by this definition of $\inherits$.
 For instance, let us call  {$0$-prime implicant} (resp. {$1$-prime implicant}) {of
$f:\Bn\to\B$}  a prime implicant of $\neg f(x)$ (resp. of $f(x)$).
 For  $\pot{t+1,i}$ to inherit from $\pot{t,j}$, we could require that $x(t)_{j\to
i}$ be involved in a \mbox{$x(t+1)_i$-prime} {implicant of $f_i$}.
This new, stronger requirement would imply the looser one of  \ref{transmission}.
But for now, we keep the looser version of \ref{transmission} because it is lighter to manipulate.

\pgh{}$\inheritss$ denotes the transitive and reflexive closure of
$\inherits$. When $\pi_1\inheritss \pi_2$ holds, we also say that $\pi_2$ inherits from $\pi_1$.

\begin{lemma}\label{lineages to paths} $\forall i,j\in V$ and $\forall t,t'\in \alltimes $, if 
 $\pot{t_1,j}\inheritss \pot{t_2,j}$  holds, then there is a path of sign $\nabla
 x(t_1)_{j}\cdot\nabla x(t_2)_{i}$ in $G$ from $j$ to $i$.
\end{lemma}

\begin{myproof}
By induction on the length of the lineage, using the definition of $\inherits$
in \ref{transmission} and the definition of equality amongst potentials in \ref{equality}.
\end{myproof}

\pgh{Potential representatives.} For any potential  $\pi$ and any time $t$, we
define 
$$\reps(\pi,t)=\{j\in
V\,:\, \pi\inheritss \pot{j,t}\}$$
as the set of automata that represent or carry $\pi$ a time $t$.
When $\pi=\pot{0,i}$ is an {original} potential, we rather write this set $\reps_i(t)$:
$$\reps_i(t)=\reps(\pot{0,i},t).$$

\pgh{Potential Charge.} 
For any automaton $i\in V$ and any time step $t$,
\begin{align*}
\orpott(i,t)=&\set{\pi:\, \pi\inherits \pot{t,i}}\\
\orpot(i,t)=&\set{\pi:\, \pi\inheritss \pot{t,i}}=\set{\pi:\, i\in \reps(\pi,t)}
\end{align*}
denotes the set of potential carried by $i$ at time $t$,
and
\begin{align*}\orpott_0(i,t)=&\set{\pot{0,j}\in \orpott(i,t)}\\
\orpot_0(i,t)=&\set{\pot{0,j}\in \orpot(i,t)}=\set{\pot{0,j}:\, i\in \reps_j(t)}
\end{align*}
denotes the set of original potential carried by $i$ at time $t$, and
$$\orpot_0(t)=\bigcup_{i\in V}\orpot_0(i,t)$$
denotes the set of original potential still represented in the BAN at
time $t$.

\begin{lemma}
\label{nonemptycharge}
From the first time $i\in V$ is updated onwards, $i$ carries potential:
$$
i=\nu(t)\implies \forall s> t: \orpott(i,s)\neq \emptyset.
$$
\end{lemma}

\begin{myproof} \ref{straightagain}.
\end{myproof}

\pgh{}\label{samepot} Note that an automaton $i$ can inherit the same potential $\pi$ several times,
at different time steps. {\color{black} In
Example~\ref{totally positive example}, automaton $1$ inherits potential
$\pot{0,5}$ at times steps $1$ and $7$:
$1\in \reps_5(1)\cap \reps_5(7)$. Indeed we have:
$\pot{0,5}\cat \pot{1,1}$ and 
$\pot{0,5}\cat \pot{1,1}\cat \pot{2,3}\cat \pot{6,4}\cat \pot{7,1}$.
\begin{lemma}
\label{nice effect}
In a nice BAN, an automaton $i\in V$  always makes the same move at times it
 inherits the same  potential $\pi$:
 $$x_i(t_1)\neq x_i(t_2)\implies i\notin \reps(\pi,t_1)\cap\reps(\pi,t_2).$$
Equivalently:
$$x(t_1)_i\neq
x(t_2)_i\implies \orpot(i,t_1)\cap\orpot(i,t_2)=\emptyset.$$
\end{lemma}

\begin{myproof} Let $\pi=\pot{t,j}$. By definition of $\reps$,
 $i\in \reps(\pi,t_1)\cap\reps(\pi,t_2)$ implies
 $\pi\inheritss \pot{t_1,i} \wedge \pi\inheritss \pot{t_2,i}$. By
 Lemma \ref{lineages to paths}, this implies there is a path of sign $\nabla
 x(t)_{j}\cdot\nabla x(t_1)_{i}$  from $j$ to $i$, as well as
 a path of sign  $\nabla
 x(t)_{j}\cdot\nabla x(t_2)_{i}$  in $G$. \textls[-10]{The BAN being nice, these paths must
 have the same sign so $\nabla x(t_1)_{i}=\nabla x(t_2)_{i}$ must hold.}
\end{myproof}

\pgh{}
Let us also note that several potentials can be transmitted to the same automaton
at once. 
In Example~\ref{totally positive example}, because
$f_1(x)=x_4\wedge x_5$, this happens to automaton $1$ every time it moves up to
state $1$ from state $0$. For instance,  at time $t=4$, it inherits the
potentials that automata $4$ and $5$ were carrying at time $t=3$:
$\pot{3,4}\inherits \pot{4,1}$ and $ \pot{3,5}\inherits \pot{4,1}$.

\pgh{} An
automaton can also inherit a  potential several times at once through different lineages.
{\color{black}This is the case with automaton $5$ and potential $\pot{0,5}$ in
Example~\ref{totally positive example}. Indeed
$\pot{0,5}\cat \pot{1,1}\cat\pot{2,3}\cat\pot{8,5}$ and
$\pot{0,5}\cat \pot{1,1}\cat\pot{2,3}\cat\pot{6,4}\cat\pot{8,5}$}.


\begin{lemma}
\label{potentials}
Each time an automaton moves, it inherits
potential that it never represented in the past. Formally, 
$\forall t\in \alltimes $, $i=\nu(t-1)$ satisfies:
 $$\orpott(i,t)\setminus \bigcup_{t'< t}\orpott(i,t')\neq \emptyset.$$
\end{lemma}

 \begin{myproof} 
Because of the equality relation among potentials, we can concentrate on time
 steps at which $i$ is updated. Let $T_i^+=\{t'< t\,: i=\nu(t'-1)\wedge 
 x(t')_i =  x(t)_i\}$ (resp. $T_i^-=\set{t'< t\,: i=\nu(t'-1)\wedge 
 x(t')_i=\neg  x(t)_i}$) be the set of dates before $t$ at which $i$ moves 
away from (resp. towards) state $x_i=x_i(t)$.
Let $t^+=\max T_i^+$.
Since $x_i(t^+)=x_i(t)$ and $i=\nu(t^+-1)=\nu(t-1)$ and $t^+<t$, there must be a time
step $t^{-}\in
T_i^-\cap\isetl{t^+,t}$.
Indeed, at both time steps $t^+-1$ and $t-1$, $i$ moves away from state $x_i$. There
 must be a time step in between at which $i$ moves towards $x_i$. And actually,
 this time step necessarily is $t^{-}-1=\max T_i^{-}-1$. 
Let $W^-_i=\set{j\in V_{\to i}:\, x_{j\to i}(t^--1)=x_i}$ be the set of
neighbours of $i$ that favour the move $i$ makes at time  $t^--1$ towards
$x_i$. If none of those neighbours change states between $t^-$ and $t$, then $i$
has no incentive to move again. But it does at time $t-1$. So one of those
neighbours $j\in W^-$ must have moved back so that $x_{j\to i}(t-1)=\neg x_i$.
Then, $\pi=\pot{t-1,j}\inherits \pot{t,i}$. And since $\pi$ is born between
$t^-$ and $t$, it cannot have been inherited by $i$ any time before $t$. 
\end{myproof}

\pgh{} Lemma~\ref{potentials} is going to be very useful because it allows to match
{injectively} each move of automaton $i$ on  $\traj$ to a path of $G$
corresponding to the lineage of a brand new potential it inherits then. 
If the BAN contains no other cycles than source loops, then this
limits the number of times an automaton needs be updated because it limits the
number of times the automaton can move.\linebreak
Precisely, in this case, Lemma~\ref{potentials} implies that no automaton $i$ can
move a number of  times that is greater than the length of the longest
loopless path going through $i$ (recall that source automata don't move at all).
\medskip
 
If there {\em are} {real} cycles in $G$ (not positive source loops) however,
Lemma~\ref{potentials} does not limit the total number of times an automaton
moves along a shortest trajectory.

\pgh{Loosing potential.} A potential $\pi$ can cease to be represented in the BAN:
$\reps(\pi,t)=\emptyset$ for some $t$. When that happens, the potential is obviously
lost for good: $\reps(\pi,t)=\emptyset\implies
\forall t'\geq t,\, \reps(\pi,t')=\emptyset$. This happens when
at some point, there is only one automaton representing potential $\pi$, and
this automaton is updated before it gets a chance to transmit its potential to
one of its out-neighbours. In  Example~\ref{totally positive
example}, at time step $0$, automaton $1$ is updated and the potential $\pot{0,1}$
is lost for good.


\pgh{Survivor potential.}
We call {survivor potential} any potential $\pi$ that is represented in the
destination configuration $y$: $\reps(\pi,T)\neq \emptyset$.  Unless
$(\nu(0),\nu(0))\in A$ is a negative loop over the first automaton that moves in
$\traj$, the potential $\pot{0,\nu(0)}$ is lost, so among the {\em original}
potentials, there are no more than $n-1$ survivors.


\begin{lemma}
\label{survivor}
 On a shortest trajectory $\traj$, at each time step, only survivor potential is
represented and transmitted.
\end{lemma}

\begin{myproof}
Let $t<T$ be the last time step before $T$ where non-survivor potential is
inherited.  Note that
$T-t=1$ is not possible unless $\nu(t-1)=\nu(T-1)$. A shortest trajectory
doesn't update twice the same automaton in a row. 
Generally, let $i=\nu(t-1)$. There necessarily exists $t-1<t'-1<T$
s.t. $i=\nu(t'-1)$. Otherwise, $i$ carries the same non-survivor potential until
the end which contradicts the definition of survivor potential.  Since no
non-survivor potential is transmitted after $t-1$, no automaton inherits the
non-survivor potential $i$ is carrying at time $t$. In other terms, $i$ could
just as well not have changed states and inherited this potential, it wouldn't
have bothered any other automaton that was supposed to change states after
$t$. A shortest trajectory wouldn't update $i$. 
\end{myproof}

\pgh{Updates and Effective updates (moves).}
From the fact that {\em once lost, an (original) potential can never be
recovered}, it is tempting to derive that recurrent configurations cannot lose
(original) potential. However, what (original) potential is present in a
configuration depends on the trajectory that lead to the configuration and not
just on this configuration itself.
Here, we need to emphasise the difference between $i$ making move $\nabla
x(t)_i$ from configuration $x(t)$, and $i\in V$ merely having its state {\em
updated} in configuration $x(t)$. When $i\in U(t)$, the update of $i$'s state is
equivalent to (/immediately results in) $i$ making move $\nabla x(t)_i$. When
$i\in S(t)$, the update of $i$'s state is ineffective (in the sense that it does
not result in $i$ moving), although still
possible. The difference this latter kind of update makes is therefore not
appreciable in the value of $i$'s state. It is, however, in terms of the
potential $i$ carries.\medskip

\fbox{\parbox{\linewidth}{
\begin{example}
\label{recurrence and potential}
{\small
$\ntk=\{f_i,\ i\in V=\{1,\ldots, 4\}\}$ where:
$$
\begin{cases}
f_1 &=~ x_1\\
f_2 &=~ x_2\\
f_3 &=~ x_1\\
f_4 &=~ (x_1\wedge x_3)\vee x_2
\end{cases}
$$

Consider the following series of updates, the first two of which define a
trajectory $\traj$ from $x=x(0)=(1,1,0,0)$ to fix point $y=x(2)=(1,1,1,1)$:\\
$
x(0)=(1,1,0,0) \stackrel{4}{\longrightarrow}
x(1)=(1,1,0,1) \stackrel{3}{\longrightarrow}
x(2)=(1,1,1,1) \stackrel{4}{\dashrightarrow} x(3)=(1,1,1,1).
$\\[3mm]
\pardessus{-0.2}{-0.2}{{\color{black!60}\scriptsize$\begin{array}{lll}\reps_1(0)&=&\set{1}\\\reps_2(0)&=&\set{2}\\\reps_3(0)&=&\set{3}\\\reps_4(0)&=&\set{4}\end{array}$}}
\pardessus{2.75}{-0.2}{{\color{black!60}\scriptsize$\begin{array}{lll}\reps_1(1)&=&\set{1,4}\\\reps_2(1)&=&\set{2,4}\\\reps_3(1)&=&\set{3,4}\\\reps_4(1)&=&\emptyset\end{array}$}}
\pardessus{5.7}{-0.2}{{\color{black!60}\scriptsize$\begin{array}{lll}\reps_1(2)&=&\set{1,3,4}\\\reps_2(2)&=&\set{2,4}\\\reps_3(2)&=&\set{4}\\\reps_4(2)&=&\emptyset\end{array}$}}
\pardessus{8.65}{-0.2}{{\color{black!60}\scriptsize$\begin{array}{lll}\reps_1(3)&=&\set{1,3,4}\\\reps_2(3)&=&\set{2,4}\\\reps_3(3)&=&\emptyset\\\reps_4(3)&=&\emptyset\end{array}$}}
\vspace{8mm}

Consider also the following series of updates, the first and only two of which also
define a trajectory $\traj'$ from configuration $x=z(0)$ to configuration $y=z(2)$:
$$
z(0)=(1,1,0,0) \stackrel{3}{\longrightarrow} z(1)=(1,1,1,0) \stackrel{4}{\longrightarrow} z(2)=(1,1,1,1)=x(2).
$$
\pardessus{0.85}{-0.2}{{\color{black!60}\scriptsize$\begin{array}{lll}\reps_1(0)&=&\set{1}\\\reps_2(0)&=&\set{2}\\\reps_3(0)&=&\set{3}\\\reps_4(0)&=&\set{4}\end{array}$}}
\pardessus{3.8}{-0.2}{{\color{black!60}\scriptsize$\begin{array}{lll}\reps_1(1)&=&\set{1,3}\\\reps_2(2)&=&\set{2}\\\reps_3(3)&=&\emptyset\\\reps_4(4)&=&\set{4}\end{array}$}}
\pardessus{6.73}{-0.2}{{\color{black!60}\scriptsize$\begin{array}{lll}\reps_1(1)&=&\set{1,3,4}\\\reps_2(2)&=&\set{2,4}\\\reps_3(3)&=&\emptyset\\\reps_4(4)&=&\emptyset\end{array}$}}
\vspace{8mm}

At the end of both series, all sets $\reps_i(t)$ have become definitely
stable. The trajectory $\traj$ embedded in the first series, however, carries
potential $\pot{0,3}$ along longer than is needed.}
\end{example}}}\medskip
 
Example~\ref{recurrence and potential} proves that a trajectory might reach a
recurrent configuration before it looses the possibility to loose potential.  It
remains the following question: {\em What configurations other than
recurrent configurations are there that have no ability to loose potential?} in
other terms: {\it Having lost the possibility to lose potential, can $\ntk$
still move far enough away from a configuration and reach a point beyond which
it can never get back to this configuration?}
\medskip

Let us extend our notations to allow for time steps at which no moves are
made but some ineffective updates are made.

\pgh{Updates and Streamlines.} We have been defining  trajectories
$\traj=(x(t))_{t\in\alltimes}$ as series of configurations starting in an initial
configuration $x=x(0)$ and ending in a final target configuration $x(T)=y$. We
could just as well have defined them with a couple $(x,(\nu(t))_{t\in\alltimes})$
comprised of the initial configuration $x$ and the series of automata $\nu(t)$
moving (being effectively updated) at each time step $t\in\alltimes $ to get from
$x(t)$ to $x(t+1)$.
With the same kind of definition we can introduce formally {\em series of
automata updates} aka {\em streamlines}: $(x,(\lambda(t))_{t\in\alltimes})$ where
$\forall t\in\alltimes ,\, \lambda(t)\in V$. At time $t+1$ of a streamline
$\Lambda=(x,(\lambda(t))_{t\in\alltimes })$, $\ntk$ is in the configuration $x(t+1)$,
where $\forall j\neq \lambda(t):\, x(t+1)_j=x(t)_j$ and automaton $i=\lambda(t)$
is in state $x(t+1)_i=f_i(x(t))=\begin{cases}x(t)_i&\text{if }i\in S(t)\\\neg
x(t)_i &\text{if }i\in U(t)\end{cases}$.
\smallskip

The transmission relation $\inherits$ extends naturally to streamlines with the
following definition: $\forall t\in \alltimes ,\ \forall j,i=\lambda(t)\in
V,\ \pot{t,j}\inherits\pot{t+1,i} \iff \sign(j,i) = \nabla x(t)_{j}\cdot
\nabla x(t+1)_{i}$.\smallskip

From now on we assume this more general definition of $\inherits$ (without loss
of anything said before using the original definition).

\pgh{Beyond.}
 Consider all streamlines whose first $T$ steps are identical to those of
trajectory $\traj$.  Among all these streamlines, consider those that are long
enough to have reached a point $\widehat{T}\geq T$ where the set of original
survivor potential is stable (\ie{} super). This must happen because, a
streamline cannot indefinitely loose original potential. It must stop loosing
original potential before it has none left.

\pgh{Super survivor potential.}
When $\traj$ leads to a recurrent configuration $y$, we call {\em super survivor
potential} any survivor potential that lingers after {\em any} series of
additional updates in $y$. In Example~\ref{recurrence and potential}, among the
set $\set{\pot{0,1},\pot{0,2},\pot{0,3},\pot{0,4}}$, of original potentials of
$\traj$, three are survivor potentials of $\traj$: $\pot{0,1},\pot{0,2}$ and
$\pot{0,3}$, while only two are {\em super} survivors of $\traj$: $\pot{0,1}$
and $\pot{0,2}$.\medskip

In a recurrent configuration, super-survivor potential can be carried  by
automata with loops over them that have the ability of keeping their charge when
they are updated. But if it is not, then super-survivor potential  must be
represented by at least two different automata.
\medskip

By Lemma~\ref{survivor}, from the very beginning, it is never useful to move an
automaton that is on the verge of inheriting non-\,surviving potential.  \forlater{We
would like to know if the same holds as well for non-{\em super}\,surviving
potential:} 

 \begin{conjecture}
\label{supersurvivor}   {\color{black!70} (If
the only cycles in $G$ are source loops, then)} from an arbitrary configuration
 $x$ to a recurrent configuration $y$, there is a shortest trajectory in which
 at each time step, only super survivor potential is transmitted.
\end{conjecture}

\forlater{
If Conjecture~\ref{supersurvivor} isn't true, then we would like to understand
why. In other terms we would like to understand what is survivor potential that
is not super survivor potential,  what is the need for it, and
where does it lie?}
\medskip

Let us point out that there can be several trajectories $\traj$ of the kind
mentioned in Conjecture~\ref{supersurvivor} leading to a recurrent configuration
$y$. An example is the case of isolated cycles. The recurrent configurations of
a cycle eventually only carry around $1$ super-survivor potential. If there is
any automaton to move in $x=x(0)$ to get to $y$, \ie{} if $x\neq y$, and if the
cycle is positive (resp. negative), then any
one of the automata that are already in their target state (resp. any one of the
$n$ automata) can serve as placeholders for the single original potential
destined to survive in $y$ and beyond.

\pgh{Attractor.} 
In the sequel, an attractor is a maximal set of recurrent configurations with
trajectories going to an back each configuration in this set.\medskip

We take interest in what we refer to as shortest trajectories between
configurations $x$ and attractors ${\cal A}\subset \Bn$,
a.k.a. ``configuration-attractor shortest trajectories''. What we mean by this
is shortest trajectories from $x$ to any of the configurations $y\in {\cal
A}$. Since the trajectories we consider are shortest trajectories, this implies
$y$ to be among the configurations of ${\cal A}$ that are the closest to $x$.

\pgh{Depths and grounds.}\label{depth} In the sequel, given a set of automata $W_0$ that we
call the {\em grounds} of $G$, we define level $d$ of $G$ as the set $W_d\subset V$
of automata whose longest path to  nodes of $W_0$ have length $d$.
The depth of automata is given by function $\omega$: $\forall i\in W_d,\, \omega(i)=d$.
Starting from an arbitrary configuration $x$, if nodes of lower depth $d>0$ can
be moved  before nodes of greater depth,
and if all non-source nodes can be moved, then the original potential initially
carried on the grounds is be survivor potential, spreading in less $n$
steps.\medskip

The proof of the next lemma is very similar to proofs given in \cite{GIGs}.

\begin{lemma}[Single paths and  cycles]\label{linear}
If $G$ is a single directed path, then:
\begin{enumerate}
\item The shortest trajectories between any two
configurations have length at most ${\cal O}(n^2)$. 
\item The shortest trajectories between any 
configurations and any  attractor have length at most $n$. 
\item On configuration-attractor  shortest trajectories, only one potential is
survivor potential: the original potential carried on the grounds by the source
automaton (the attractor is a stable configuration). And each automaton whose
original state differs from that of the source automaton is updated exactly
once. Other automata are not updated.
\end{enumerate}
If $G$ is a positive cycle, then:
 \begin{enumerate}[resume]
\item All the same holds. In Item 3, on the grounds, ``source automaton'' must be replaced by
 ``any automaton whose initial state is already equal to its final state''.
\end{enumerate}
If $G$ is a negative cycle then:
 \begin{enumerate}[resume]
\item Again, Items 1, 2 and 3 still hold. 
 In Item 3 ``source automaton'' must be replaced by
 ``any automaton''.
\item From an arbitrary configuration, all recurrent configurations can be
reached in at most $2n$ steps.
\end{enumerate}
\end{lemma}

\begin{myproof}
Item 3 is immediate. Item 2 is the immediate consequence of Item 3. Item 1
follows from the following. There are at most $n$ original potential that
survives. None of them can get inherited more than once by the same automaton.
Each of them can be inherited by at most $n$ automata. Item 4 comes from the
fact that positive cycles behave just like single paths except for the following
difference. In a positive cycle, any automaton can be the one that imposes its
original potential to the others. And actually, if two automata are connected by
a positive path, then eventually it will not matter which of them is chosen. A
positive cycle has two alternatives. A single path has only one. Other than that
both types of structures behave exactly the same way on shortest trajectories.
\end{myproof}

All cases of Lemma~\ref{linear} can be seen in terms of dropping original
potential down a single path.  In all cases, no potential that eventually
disappears needs ever be transmitted.
And informally, if an original potential travels all around the path, then
either that is because this potential has ridden the whole BAN from all other
original potential. Or this potential and others have been cycling unnecessarily
around the cycle. 
There are only two cases in which the cyclic nature of the BAN's structure
really counts: in the case of a positive cycle, in the choice mentioned in the
proof above, and in the case of a negative cycle, in the cyclic
attractor.\medskip

\forlater{{\bf Important Remark:} This example suggests that on the way to a recurrent configuration, a potential has no need to go twice through a path that transmits it
without transformation.}\smallskip

\begin{lemma}[Acyclic $G$]\label{Gacyclic}
If $G$ is acyclic (except for the source loops), then  shortest trajectories to
attractors have length at most $n$. And in this case, there is only one
attractor which is a stable configuration.
\end{lemma}

\begin{myproof}Define the grounds as the set of source automata, and \cf \ref{depth}.
\end{myproof}


\pgh{}In the sequel,  $\nabla_i=-\nabla y_i$ denotes a
move {\em towards} $y$, the ultimate move that $i$ must have made (or at least
not un-made) on $\traj$.

\pgh{(Dis)Favourable Potential.} First, we say that a potential $\pi=\pot{t,j}$ is {favourable} (resp. {disfavourable}) to automaton $i$ or to move $\nabla_i$ if
all paths from $j$ to $i$ have sign $\sign^\ast(j,i)=\nabla
x(t)_j\cdot \nabla_i$ 
(resp. $\sign^\ast(j,i)=-\nabla
x(t)_j\cdot \nabla_i$).  Nice BANs are the ones where $\forall \pi,\forall i$,
$\pi$ is either favourable or disfavourable to $i$. \linebreak
A potential $\pi$ that is neither favourable nor disfavourable to $i$ is said to
be undecided to $i$. Nice BANs have no undecided potential.

\begin{lemma}\label{favorable survivor potential}
In $y$, an automaton $i$ only represents survivor potential that is favourable to $i$.
\end{lemma}

\begin{myproof}By definition of $\inherits$ and of favourable.
\end{myproof}

\pgh{(Dis)Favourable Neighbours.} 
We introduce the following two sets relative to trajectory $\traj$:
$$
\begin{cases}
\Aplus&=~\{(j,i)\in A\ :\ \sign(j,i)=+\nabla_{j}\nabla_{i}\}\\
\Amoins &=~\{(j,i)\in A\ :\ \sign(j,i)=-\nabla_{j}\nabla_{i}\}.
\end{cases}
$$
In-neighbours $j:\,(i,j)\in\Aplus$ (resp. $\in \Amoins$) are called favourable
(resp. disfavourable) in-neighbours of $i$. Them having already (resp. them
having not yet) made move $\nabla_j$  and already being in state $y_j$
(resp. still being in state $\neg y_j$) can only favour $i$ in making move
$\nabla_i$. The following Lemma is thus about the most favourable conditions for
$i$ to make move $\nabla_i$.

\begin{lemma} \label{most favorable}
\mbox{$\forall i\in V$,  $\forall z\in \Bn$:
$
\begin{cases}z_j=y_j, &\forall j:\,(i,j)\in\Aplus \\ 
 z_j=\neg y_j,&\forall j:\,(i,j)\in\Amoins
\end{cases}
~\implies ~i\in U(z).
$
}\end{lemma}

Note that a disfavourable neighbour acts favourably when it is in state $\neg
y_k$. And a favourable neighbour acts disfavourable when it is not in state
$y_j$.

\pgh{Cycles and disfavours.} Note also that it is not possible to have cycles in
$G$ comprised solely of arcs
in $\Amoins$. And  a cycle comprised of an even  (resp. odd) number of
arcs in $\Amoins$ is necessarily a positive (resp. a negative) cycle .

\begin{lemma}Assume that $\forall i\in V$, $i$ either favours all his out-neighbours, or
  disfavours them all.  Then, $\traj$ (a shortest
  trajectory) has length at most $n$.
\end{lemma}

\begin{myproof}
As long as there are unstable automata of the first kind (that favour all their
out-neighbours) that are in state $\neg y_i$, those can be moved without adding
any disfavour to any move $\nabla_j$ that still needs to be made.  When we get
to the point where all unstable automata that need to move are of the second
kind (disfavouring all their out-neighbours), all their out-neighbours
necessarily already are in state $y_j$.
\end{myproof}

\begin{lemma}\label{no A moins}\label{Hcyclicplus}If $\Amoins=\emptyset$ and $\traj$   is a shortest
trajectory, then $\traj$  (a shortest trajectory) has length at most $n$. And in this case, if $\traj$ is
a shortest trajectory to an attractor, the attractor is a stable configuration,
one of at most two if $G$ is strongly connected ({\it eg} if $G$ is a positive
cycle).
\end{lemma}

\begin{myproof}No automaton $i$ has disfavourable neighbours.
No move $\nabla_j$ can disfavour $\nabla_i$. And no move $-\nabla_j$ will favour
any move  $\nabla_i$. So only moves $\nabla_i$ need be made. And all $\nabla_i$
moves {\em can} be made.\medskip

The second part of Lemma~\ref{no A moins} comes from the fact that once all
 automata $i$ have made move $\nabla_i$, none of them can make move $-\nabla_i$
 since none of them has in-neighbours favourable to that move.
\end{myproof}

\pgh{Favour Graph.}
We define graph $H^{\traj}=(V,\AH)$ where $\AH=\Aplus\cup\set{(i,j):\, (j,i)\in
\Amoins}$. 



\begin{lemma}\label{Hacyclicbary}If  $H^{\traj}$ is acyclic except for loops,
and if no automata $i$ are initially already in their destination state $y_i$,
except possibly the source automata,
then $\traj$, a shortest trajectory from $x$ to $y$ has length less than $n$.
\end{lemma}

\begin{myproof}We let the grounds  of $H^{\traj}$ be  the set of source
automata.  We let $\mu:V\to \iset{1,n}$ be an injective function satisfying
$\forall (j,i)\in\AH, i\neq j$, $\mu(j)<\mu(i)$ and $\forall i,j\in
V,\, \omega(j)<\omega(i)\implies \mu(j)<\mu(i)$. This function exists because
$H^{\traj}$ is acyclic.
We show that $\forall t,\, i=\mu^-1(t)\in U(t)$ by induction on $t$, by showing
that the conditions of Lemma \ref{most favorable} are maintained. Thus we can
let $\nu=\mu^-1$ so that at each time step $t<n$ the non-source automaton
$\nu(t)$ is updated.
\end{myproof}

\begin{lemma}\label{HacyclicFP}If  $H^{\traj}$ is acyclic except for  loops, and if
$y$ is stable configuration $\traj$, a shortest trajectory from $x$ to $y$ has
length less than $n$.
\end{lemma}

\begin{myproof}We apply the same process as in the proof of
Lemma \ref{Hacyclicbary}: we define the source automata to be the grounds, and  move automata from shallowest to deepest.
However, in this case does the process does not guarantee maintaining the ideal
conditions of Lemma \ref{most favorable} because some disfavourable in-neighbours
$k$ of $i$ might already be in state $y_k=x_k$.
Lemma \ref{HacyclicFP} is proven by comparing the inputs of $i$ on an arbitrary
configuration $x(t)$ of $\traj$, and the inputs of $i$ in the stable
configuration $y$. In both $x(t)$ and $y$, all favourable in-neighbours of $i$ are
in their favourable state (in $x(t)$ this is because of the order according to
which we are moving automata). The difference between the two situations is that in $x(t)$,
not all disfavourable neighbours of $i$ are in their disfavourable state.
Thus, if $f_i(y)=y_i=\neg x_i(t)$ (and it is), then $f_i(x(t))=y_i=\neg x_i(t)$
and $i$ can indeed be moved in that configuration.
\end{myproof}

{In Lemma \ref{HacyclicFP}, $\Htraj$ is acyclic. $G$ isn't
necessarily. The source automata of $\Htraj$ aren't necessarily source automata
 of $G$.  {\em Something} must be keeping them in their state $y_i$. Since they
 are source automata of $\Htraj$, all their in-neighbours distinct from
 themselves are disfavourable. Thus, there must be a positive loop over each one
 of them.}

\begin{lemma}
\label{HacyclicOscil}If  $H^{\traj}$ is acyclic except for  loops ({\it eg} if $G$
 is a negative cycle), then $\traj$ a shortest trajectory from $x$ to an
 attractor ${\cal A}$ has length no greater than $n$.
\end{lemma}

\begin{myproof}Here, we apply the same process as before:  we define the source
 automata of $H^{\traj}$  to be the grounds, and  move automata from shallowest to deepest.
However, here again, we cannot guarantee the ideal conditions of Lemma \ref{most
favorable} as in Lemma \ref{Hacyclicbary}. Nor can we rely on the stability of
the destination configuration $y$ as in Lemma \ref{HacyclicFP}.
Instead we are going to exploit the following flexibility we have. $\traj$ needs
to end on a configuration $y$ of ${\cal A}$, and {\em any} configuration
$y\in{\cal A}$ will do.
We will exploit this flexibility by allowing ourselves to change our target
configuration $y$ along the way.
More precisely, we initialise our target to a certain $y=y(0)\in {\cal A}$.
 Then, at each time step we  update our target if we know of a closer
 $y(t)\in {\cal A}$ belonging to the same attractor and that can be reached from
 $x(t)$.
The idea is to maintain at each time step $t$, the property that the automaton
$i=\nu(t)$
that we are considering for update has all its favourable in-neighbours $j$ in
state $y_j(t)$.  
So we want to ensure that \litemr $\forall t$, $x=x(0)$ can indeed reach
$x(t)$, and \litemr\endlitemr either $HD(x(t),y(t))>HD(x(t+1),y(t+1))$, or no
move is made at time $t$  in which case $HD(x(t),y(t))=HD(x(t+1),y(t+1))$ and
step $t$ of the process doesn't count as a step of the trajectory.\smallskip

By definition, the source automata $i$ of $\Htraj$ have no favourable 
in-neighbours $j\neq i$, and thus  no favourable 
in-neighbours that aren't already in state $y(0)_j$. 
\smallskip

Let $t$ be an arbitrary time step of the process where $i=\nu(t)$ is
considered. There are three cases:
\begin{enumerate}
\item $i\in U(t)$ and $x(t)_i= y(t)_i$. In this case, $i$ is not moved we move
on to $t+1$ and to automaton $\nu(t+1)$.
\item $i\in U(t)$ and $x(t)_i\neq y(t)_i$. In this case,  $i$ is updated to state $y_i$. We let 
$x(t+1)=\bar{x(t)}{i}$ be the configuration that is reached from $x(t)$ by
moving $i$, and we maintain the target $y(t+1)=y(t)$, so that
$HD(x(t+1),y(t+1))= HD(x(t),y(t))-1$.
\item $i\in S(t)$ and $x(t)_i\neq y(t)_i$. In this case, we cannot move $i$ to
get closer to our current target. So we let $x(t+1)=x(t)$ and change our target
to $y(t+1)=\bar{y(t)}{i}$ because of the following reasons.
In $y(t)$, $i$ must be unstable ($i\in U(y(t))$). Indeed, in $y(t)$, all
  automata $j$ are in state $y(t)_j$. In particular, all in-neighbours $j$ of
  $i$ favouring $i$ being in state $y(t)_i$ are in state $y(t)_j$. They also
  already are in $x(t)$. As far as $i$ is concerned, the difference between
  $x(t)$ and $y(t)$ is that in $x(t)$ there are less disfavourable in-neighbours
  presently disfavouring $i$ being in state $y(t)_i$. Yet $i$ still cannot move
  to state $y(t)_i$: $f_i(x(t))=x(t)_i=\neg y_i(t)$. This implies that in
  $y(t)$, the situation is even less favourable and $i$ cannot maintain state
  $y(t)_i$: $f_i(y(t))=x(t)_i=\neg y(t)_i$. 
Thus $y(t+1)$  is the configuration that is reached from $y(t)$ by moving the
unstable automaton $i$.  It therefore is also a recurrent configuration
belonging to the same attractor as the one $y(t)$ belongs to.
And we have: $HD(x(t+1),y(t+1))=
HD(x(t),y(t))-1$.
\end{enumerate}
\end{myproof}

The next lemma is redundant. Note that a nice BAN only has stable configurations
as attractors since it contains no negative cycles.

\begin{lemma}\label{Hacyclicplus} If $\ntk$ is nice, $H^{\traj}$ is acyclic except for strongly connected components comprised
 solely of arcs of $\Aplus$ then 
$\traj$ a shortest trajectory from $x$ to a stable configuration $y$ has length no greater than $n$.
\end{lemma}

\begin{myproof}
 Let $\underline{\Htraj}$ be the reduced version of
$\Htraj$, identical to $\Htraj$ except that all strongly connected components 
(SCCs) have been reduced to a single node $c\in\underline{V}$ with a loop over it. In this
case, the loop can be considered as a favouring loop:
$(c,c)\in\underline{\Aplus}$.  
The sets $\underline{\Aplus}$ and $\underline{\Amoins}$ can be defined
straightforwardly so that
$\underline{\AH}=\underline{\Aplus}\cup\underline{\Amoins}$.  Indeed, because
$\ntk$ is nice, and because of the assumption on the arcs of SCCs, the following
holds. If $j$ and $k$ are two in-neighbours (resp. out-neighbours) of
$i$ belonging to the same SCC of $\Htraj$, then they either both are favourable
to $i$ or they are both are disfavourable (resp. $i$ is either favourable to both
or disfavourable to both).
Thus, we can still order nodes of $\underline{\Htraj}$ as in the proofs of the
previous lemma in order to move them from shallowest to deepest.
To extend this order to nodes of $\Htraj$, we use Lemma \ref{Hcyclicplus} for
nodes of SCCs of $\Htraj$. 
%
This way, the automata belonging to source SCCs of $\Htraj$ are moved under the
 same conditions as the isolated SCCs  are in the proof of Lemma \ref{Hcyclicplus}.
Next come the automata that have in-neighbours in these source-SCCs.   For them, everything happens as
in the proof of Lemma \ref{HacyclicOscil}.
And when comes the time to update a non-source SCC, then again our extended
ordering of automata guarantees we can make the targeted moves as in the proof
of Lemma
\ref{Hcyclicplus} because 
 the in-neighbourhood of the SCC is consistently favourable to every automaton in
it.
\end{myproof}

\renewcommand{\refname}{}
\vspace{-5mm}

{\small \bibliography{arxiv_shortesttrajectories}}

\end{document}